\documentclass[reprint,floatfix,superscriptaddress, amsmath,amssymb]{revtex4-2}
\usepackage{graphicx}
\usepackage{amsmath}
\usepackage{color}
\usepackage{bm}
\usepackage{braket}
\usepackage[export]{adjustbox}
\usepackage[colorlinks=true,allcolors=blue]{hyperref}
\usepackage{dcolumn}
\usepackage{bm}
\usepackage{tikz}
\usepackage[caption=false]{subfig}
\begin{document}
\newenvironment{captivy}[1]{
  \begin{tikzpicture}[every node/.style={inner sep=0}]
    \node[anchor=south west,inner sep=0] (image) at (0,0) {#1};
    \begin{scope}[x={(image.south east)},y={(image.north west)}]
}
{
        \end{scope}
  \pgfresetboundingbox
  \path[use as bounding box] (image.south west) rectangle (image.north east);
  \end{tikzpicture}%
}
\newcommand*{\oversubcaption}[3]{
  \draw (#1) node[fill=white,inner sep=0pt, opacity=0.2, above, yscale=1.1, xscale=1.1] {\phantom{(a)#2}};
  \draw (#1) node[inner sep=0pt, above]{%
    \subfloat[#2\label{#3}]{\phantom{(a)}}
  };
}
\preprint{APS/123-QED}
\title{Confinement induced three-dimensional trajectories of microswimmers in rectangular channels}
\def\afflux{ Department of Chemical Engineering\\
 Indian Institute of Technology, Madras
,600036, India}
\author{Byjesh N. Radhakrishnan }%
\affiliation{\afflux}
\affiliation{Department of Physics and Materials Science,\\ University of Luxembourg, L-1511 Luxembourg, Luxembourg}
\author{Ahana Purushothaman}%
\affiliation{
 Department of Chemical Engineering\\
 College of Engineering, King Faisal University, Al-Ahsa 31982, Saudi Arabia}
\author{Ranabir Dey}
\affiliation{% 
Department of Mechanical and Aerospace Engineering,\\
Indian Institute of Technology Hyderabad,
Kandi, Sangareddy 502285, India }
\author{Sumesh P Thampi}%
\affiliation{\afflux}%
\date{\today}
\begin{abstract}
We study the trajectories of a model microorganism inside three-dimensional channels with square and rectangular cross-sections. 
Using (i) numerical simulations based on lattice-Boltzmann method, and (ii)  analytical expressions using far-field hydrodynamic approximations and method of images we systematically investigate the role of the strength and finite-size of the squirmer, confinement dimensions, and initial conditions in determining the three dimensional trajectories of microswimmers. 
Our results indicate that the hydrodynamic interactions with the confining walls of the channel significantly affect the swimming speed and trajectory of the model microswimmer. 
Specifically, pullers always display sliding motion inside the channel: weak pullers slide through the channel centerline, while strong pullers slide through a path close to any of the walls. 
Pushers generally follow helical motion in a square channel. Unlike pullers and pushers, the trajectories of neutral swimmers are not easy to generalize, and are sensitive to the initial conditions. 
Despite this diversity in the trajectories, the far-field expressions capture the essential features of channel-confined swimmers. 
Finally, we propose a method based on the principle of superposition to understand the origin of the three-dimensional trajectories of channel confined swimmers.  
Such construction allows us to predict and justify the origin of apparently complex 3D trajectories generated by different types of swimmers in channels with square and rectangular cross sections.
\end{abstract}%According to this method, the three-dimensional trajectories can be constructed from two planar trajectories, each produced by the swimmer when it is restricted to move in the center planes of the channel.
\maketitle
\section{\label{sec:level1}Introduction}
Microorganisms swim in complex natural and biological environment overwhelmingly constituted by confined spaces \citep{livingfluids, conrad2018confined}. 
A few examples are sperm cells swimming inside the female reproductive tract \citep{sperm, spermmotion}, and bacteria swimming in the digestive gut and in pores in the extracellular matrix \citep{bacteria, montecucco2001living}.
It is interesting to note that the swimming dynamics of microorganisms is determined by the stimuli or cues that they receive from the local environment, which may emanate from confining surfaces in immediate proximity and from other active or passive entities in the vicinity \citep{crowdedcells,crowdedenvironment,ahana_activepassive}. In addition to the steric \citep{steric}, chemical \citep{chemical}, electrical \citep{electric1,electric2}, and thermal \citep{thermal} interactions between a microswimmer and a confining surface, hydrodynamic interactions with the confining walls can play a dominant role in dictating the microswimmer dynamics.

 In one of the earliest experiments, it was shown that wall induced hydrodynamic interaction led to the preferential accumulation  of bull sperm cells near the confining walls \citep{Rothschild}. 
 The hydrodynamic interaction of a microswimmer that swims using flagella or a flagellum at the posterior, known as pushers (e.g. sperm cells and flagellated bacteria), with a rigid wall results in reorientation of the swimmer parallel to the wall and in a simultaneous attraction towards the closest wall  \citep{Rothschild,lauga2009hydrodynamics}.
 Subsequently, several studies focused on this hydrodynamic origin of the accumulation of swimming cells near a confining wall using both sperm cells \citep{spermatozoaatsurfaces,hydrodynamicsperm} and bacterial cells \citep{biofilm1,bacterianearsurface,Berk}. 
 Interestingly, it was also shown that the swimming trajectory of a pusher-type microswimmer turns from straight, with intermittent tumbling, in the bulk to circular close to a wall due to the involved hydrodynamic interactions \citep{lauga2006swimming,lauga2009hydrodynamics}.
 The directionality of the circular trajectory, i.e. clockwise or anti-clockwise, depends on whether the confining surface is a rigid wall or an air-water interface \citep{lauga2006swimming,lauga2009hydrodynamics,di2011swimming}. 
 Furthermore, microswimmers exhibit other surprising behaviours like orbiting/sliding, hovering, scattering, bouncing and oscillation near solid boundaries, specifically curved ones, and in confinements due to the involved hydrodynamic interactions with walls \citep{spagnolie2015geometric, lintuvuori2016hydrodynamic, Kuron2019, ahana2019confinement,chaithanya2021wall}.
 Instead of smooth and rigid walls, hydrodynamic interactions of microswimmers were also studied near corrugated periodic surfaces  \citep{periodicsurface,periodicconfinement} and elastic boundaries \citep{elasticboundary,julia}. 
 Further, the changes in the trajectories of microswimmers were studied by varying the channel geometry \citep{geometry, dhar2020hydrodynamics}, the channel size \citep{capillary,Elgeti,speedenhance}, and by including the higher-order hydrodynamic moments describing the microswimmer \citep{lowestmode}. 
 Most of these studies indicate two things- (i) confinement enhances swimming velocity, and (ii) oscillatory or helical trajectories are commonly observed when the microswimmers are confined in narrow channels or capillary tubes. A detailed numerical study \citep{capillary} of the trajectories of a single spherical squirmer in a circular tube revealed that in the cylindrical confinement neutral microswimmers adopt helical trajectories, while weak and strong pullers exhibit sliding motion through the center of the tube and sliding near the wall respectively. Pusher-type swimmers end up crashing against the walls \citep{capillary}.  Despite the importance of hydrodynamic interaction with the confining walls, studies detailing the role of geometry of the channel, in particular the role of anisotropy of the channel cross section, on the dynamics of microswimmers remain scarce. %Specifically, the dynamics of a microswimmer in microchannels with rectangular and square cross sections still remain poorly understood.

Recently, understanding the motility and collective behaviour of biological microswimmers like bacteria in microfluidic environment has gained importance for its significance in ecology, biological processes, diagnostics and bio-computation \citep{jeanneret2019confinement}.
%jetokarova2021patterns
Additionally, synthetic microswimmers that are engineered to mimic the behavior of their biological counterparts, also find applications in microfluidic devices and confinements \citep{microfluidic}.
Although microchannels with rectangular cross-sections are most common in all microfluidic applications, the role of anisotropic channel cross-section on microswimmer dynamics, and the resulting three-dimensional complex trajectories of the microswimmers remain largely unaddressed in the literature. 
The present work aims to address this lack of understanding by analyzing the trajectories of squirmers of varying strength in a range of microchannels of rectangular cross-section.

We use a spherical squirmer model for the micro-swimmer which significantly reduces the complexity associated with the shape and the swimming strokes of the microswimmer. 
We also restrict our analysis to flat-walled microchannels in order to avoid the  complexity of microswimmer dynamics near curved surfaces \citep{curvedsurface, chaithanya2021wall}. 
We study the dynamics of squirmers in these rectangular microchannels using numerical simulations which employ lattice-Boltzmann method (LBM). 
Interestingly, we also develop an approximate analytical approach for calculating the trajectories of the squirmers based on the far-field hydrodynamic approximation \citep{pozrikidis_1992,chwang_wu_1975,farfieldaccuracy,pak2014theoretical} of the microswimmer flow field along with method of images \citep{methodofimage,spagnolie_lauga_2012}. 
The results from this analytical approach are compared with those from full numerical simulations in order to test the efficacy of the former.
Our results illustrate the consequences of a rectangular confinement on the trajectory of a microswimmer in a microchannel. 
Importantly, our analyses also help in disentangling the complexity of the three-dimensional trajectories of channel confined swimmers by constructing them as superposition of trajectories from simpler configurations.

The rest of the paper is structured as follows - in section~\ref{sec:mathematical model}, the squirmer model and the geometry of the channel are described. The section \ref{sec:method} is dedicated to explaining the solution methodology wherein we discuss the method of far-field approximation along with the method of images and the lattice-Boltzmann method (LBM). We then present the results in section~\ref{sec:results} by first finding the instantaneous swimming velocity of the squirmer and then validating the accuracy of the far-field approximation. 
Then, in ~\ref{subsec:2d_results}, we move on to explore the two-dimensional (2D) trajectories of squirmers in the square channels. We will show that swimmers either swim parallel to the channel axis or perform wavelike motion depending upon the strength of the force dipole. The results are compared with far-field approximations wherever possible. The analysis is then generalised for the case of three-dimensional (3D) motion in both square (\ref{subsec:3d_trajectory}) and rectangular (\ref{subsec:3d_rectangle}) channels, and the origin of the complexity of the trajectories is explained using a simple superposition method. 

\section{Mathematical model} 
\label{sec:mathematical model}
\begin{figure*}[!tp]
    \centering
  \begin{captivy}{\includegraphics[clip,trim=0cm 0cm 0cm 0cm,width=.4\textwidth]{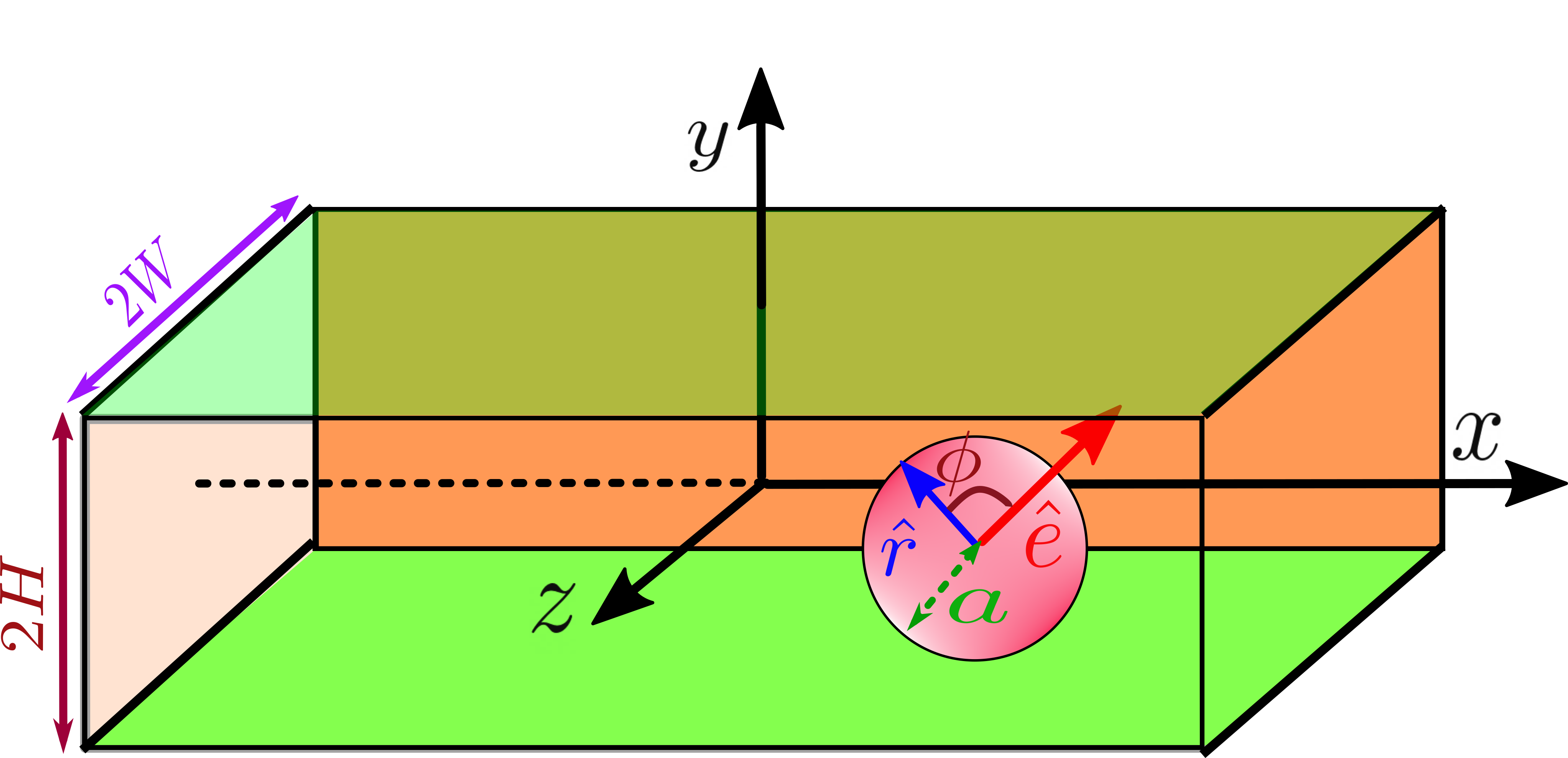}
  \hspace{1cm}
  \includegraphics[clip,trim=0cm 0cm 0cm 0cm,width=.27\textwidth]{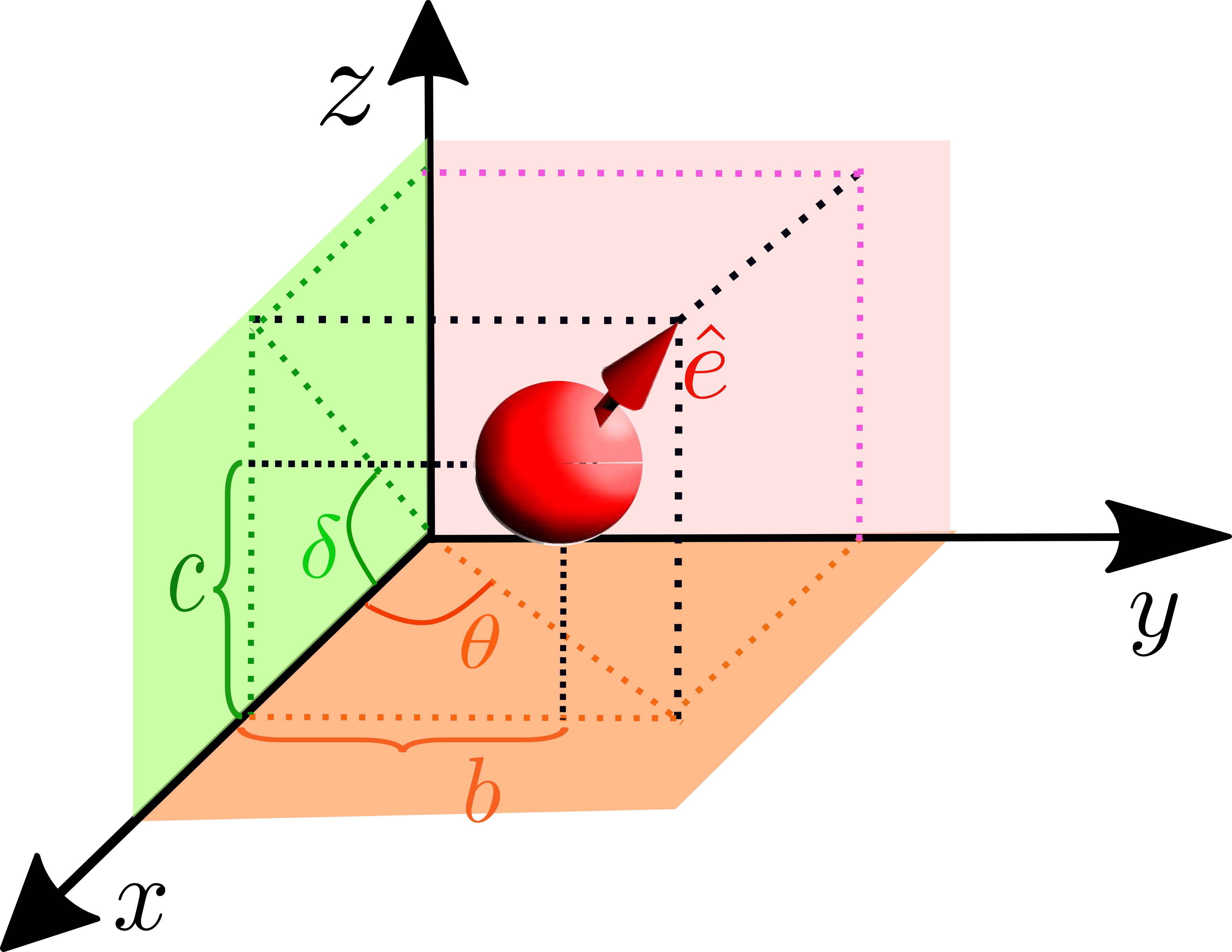}}
    \oversubcaption{0.1, 0.9}{}{fig:schem1}
    \oversubcaption{0.6, 0.9}{}{fig:schem2}
  \end{captivy}
\caption{(a) Schematic representation of a spherical squirmer of radius $a$ swimming in a rectangular channel. The unit vector $\hat{\boldsymbol{e}}$ represents the orientation of the squirmer and $\hat{\boldsymbol{r}}$ is a position vector that describes the surface of the squirmer with respect to its center. The channel is of height $2H$, and width $2W$. The origin of the Cartesian coordinates coincides with the center of the channel. The $x$-direction is periodic, and it aligns with the channel centerline, which is shown as the dashed line. The $y$ and $z$-axes  are oriented along the height and width of the channel respectively. (b) Symbols used to describe the location and orientation of the squirmer. From the channel centerline, the squirmer is located at a distance $b$ in the $y$-direction and $c$ in the $z$-direction. The angles $\theta$ and $\delta$ describe the orientation of the microswimmer, and are measured from $x$-axis to the projection of the orientation vector $\hat{\boldsymbol{e}}$ on $xy$ and $xz$ planes respectively.
} 
\label{fig:schematic}
\end{figure*}
%
%
%%%%%%%%%%%%%%%%%%%%%%%%%%%%%%%%%%%%%%%%%%%%%%%%%%%%%%%%%%%%%%%%%%%%%%%%%%%%%%%%%%%%%%%%%%%%%%%%%%%%%%%%%%%%%%%%%%%%%
\subsection{Squirmer model}
In this study, we use the squirmer model to simulate microswimmers \citep{Lighthill,Blake71,pedley}. We consider an axisymmetric, non-deformable, spherical squirmer \citep{pedley} with a tangential surface slip velocity $U_{s}(\phi)$ given by,
\begin{align}
% \boldsymbol{U}_{s}(\phi) =
U_{s}(\phi) = \sum_{n=1}^{\infty}B_{n}\frac{2}{n(n+1)}\sin (\phi) P_{n}^{'}(\cos \phi)
\label{eq:slip velocity}
\end{align}
where $B_{n}$ is the strength of $n^{th}$ mode of tangential slip velocity, and $P_{n}^{'}$ is the associated Legendre polynomial. The polar angle between the position vector on the surface of the squirmer $\hat{\boldsymbol{r}}$ and the swimming direction $\hat{\boldsymbol{e}}$ is represented as $\phi =\cos ^{-1}(\hat{\boldsymbol{e}} \cdot \hat{\boldsymbol{r}} /a)$. The swimming speed of the squirmer in an unconfined environment, $\hat{\boldsymbol{U}}_{f}=(2/3)B_{1}\hat{\boldsymbol{e}}$, is dependent on the first swimming mode alone.

In this work, we have $B_{n}=0$, for $n>2$. The relative strengths of the first two modes, $\beta=B_{2}/B_{1}$, determines the hydrodynamic nature of the swimming motion. On one hand, the squirmers with $\beta < 0$ are called pushers (e.g. \textit{Escherichia coli}). They swim by pushing the fluid at the front and back, and pulling the fluid from the sides. On the other hand, pullers, $\beta > 0$, (e.g. \textit{Chlamydomonas}) swim by pulling the fluid from front and back. The squirming motion with $\beta=0$ corresponds to neutral swimmers (e.g. \textit{Paramecium}). These microswimmers move the fluid from front to back by generating flow field corresponding to a source dipole \citep{pedley}.

%%%%%%%%%%%%%%%%%%%%%%%%%%%%%%%%%%%%%%%%%%%%%%%%%%%%%%%%%%%%%%%%%%%%%%%%%%%%%%%%%%%%%%%%%%%%%%%%%%%%%%%%%%%%%%%%%%%%%
\subsection{Swimming through a rectangular channel}
\label{subsec:flowing in a rectangular channel}
We consider the spherical squirmer of radius $a$ swimming in the rectangular channel as illustrated in Fig.~\ref{fig:schematic}. The $x$-axis is along the channel centerline, and $y$ and $z$ axes are considered along the height and width of the channel respectively. From the channel centerline ($x$-axis), the squirmer is located at a distance $b$ in the $y$ direction and at a distance $c$ in the $z$ direction. The dimensionless positions $x^{*}$, $y^{*}$, and $z^{*}$ are defined as,
\begin{align}
x^{*} & =x/a\\
y^{*} & =b/(H-a)\\
z^{*} & =c/(W-a)
\label{eq:zstar}
\end{align}
In other words, $y^{*}=0$ represents a plane passing through the channel center and perpendicular to the $y$-axis. $y^{*}=\pm1$ represent the location of the squirmer when it touches the walls perpendicular to the $y$-axis (henceforth referred to as top and bottom walls). Similarly, $z^{*}=0$ is the plane passing through the channel center and perpendicular to the $z$-axis. $z^{*}=\pm1$ represent the location of the squirmer when it touches the walls perpendicular to $z$-axis (henceforth referred to as left and right walls).

The strength or the extent of the confinement is quantified by the ratio of the squirmer radius to the half channel height, $a/H$. The aspect ratio of the channel cross-section, $AR$, is defined as
\begin{align}
AR=(W-H)/W
\label{eq:ar}
\end{align}
such that $AR=0$ represents a square channel. The rectangular channels have $AR\ne0$. $AR\xrightarrow{}1$ represents a rectangular channel with a very large width.

%%%%%%%%%%%%%%%%%%%%%%%%%%%%%%%%%%%%%%%%%%%%%%%%%%%%%%%%%%%%%%%%%%%%%%%%%%%%%%%%%%%%%%%%%%%%%%%%%%%%%%%%%%%%%%%%%%%%%
\subsection{Governing equations for the squirmer velocity field}
\label{subsec:governing equation-fluid}
The squirmer is suspended in an incompressible, Newtonian fluid, and the velocity field generated by the squirmer is described by the Navier-Stokes equations, 
\begin{align}
\nabla \cdot \hat{\boldsymbol{u}} &= 0\label{eq:incompressibility}\\
\rho \left( \frac{\partial \hat{\boldsymbol{u}}}{\partial t} + \hat{\boldsymbol{u}} \cdot \nabla \hat{\boldsymbol{u}} \right) & = -\nabla p + \mu \nabla^{2} \hat{\boldsymbol{u}}
\label{eq:NS}
\end{align}
where $\hat{\boldsymbol{u}}$ is the velocity field, $p$ is the pressure field,  and $\mu$ and $\rho$ are the viscosity and density of the surrounding liquid respectively. The periodic boundary condition is applied at the two open boundaries (along the $x$-axis) of the rectangular channel, and the no-slip boundary condition is specified at the four confining walls.

%%%%%%%%%%%%%%%%%%%%%%%%%%%%%%%%%%%%%%%%%%%%%%%%%%%%%%%%%%%%%%%%%%%%%%%%%%%%%%%%%%%%%%%%%%%%%%%%%%%%%%%%%%%%%%%%%%%%

\section{Solution methodology}
\label{sec:method}%
We use two different approaches to determine the velocities and trajectories of the microswimmer. 
The first method is an approximate analytical approach, where we use the far-field hydrodynamic approximation and the method of images to account for the hydrodynamic interactions between the microswimmer and all the channel walls. In the second method, which is a numerical method, we perform the full-scale simulation of the squirmer in a rectangular channel using LBM. These approaches are discussed below.

%%%%%%%%%%%%%%%%%%%%%%%%%%%%%%%%%%%%%%%%%%%%%%%%%%%%%%%%%%%%%%%%%%%%%%%%%%%%%%%%%%%%%%%%%%%%%%%%%%%%%%%%%%%%%%%%%%%%%
\subsection{Method 1: Far-field hydrodynamic approximation}
\label{sec:far-field}
\begin{figure*}[!tp]
    \centering

  \begin{captivy}{\includegraphics[clip,trim=0cm 0cm 0cm 0cm,width=.45\textwidth]{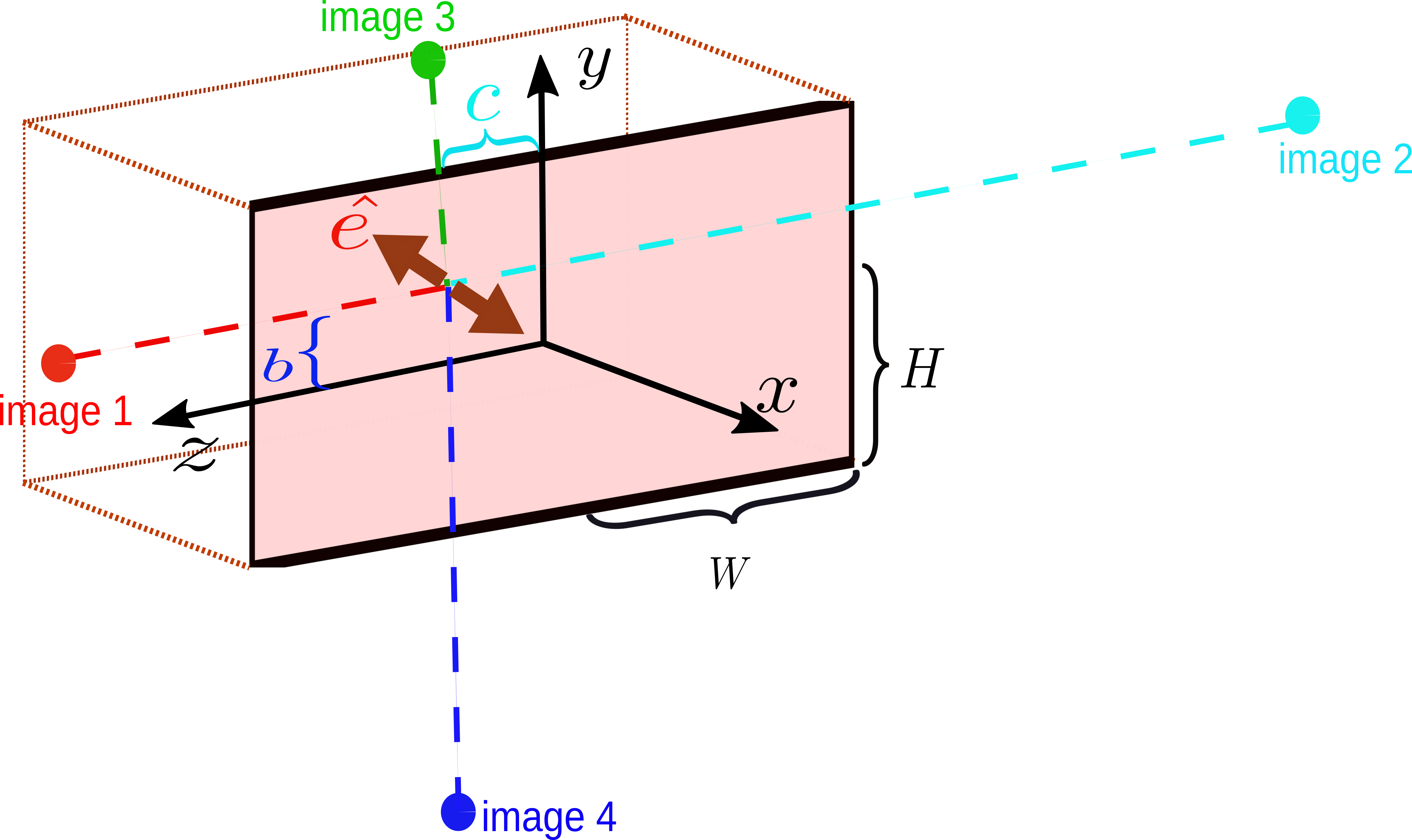}
  \hspace{1cm}
  \includegraphics[clip,trim=0cm 0cm 0cm 0cm,width=.45\textwidth]{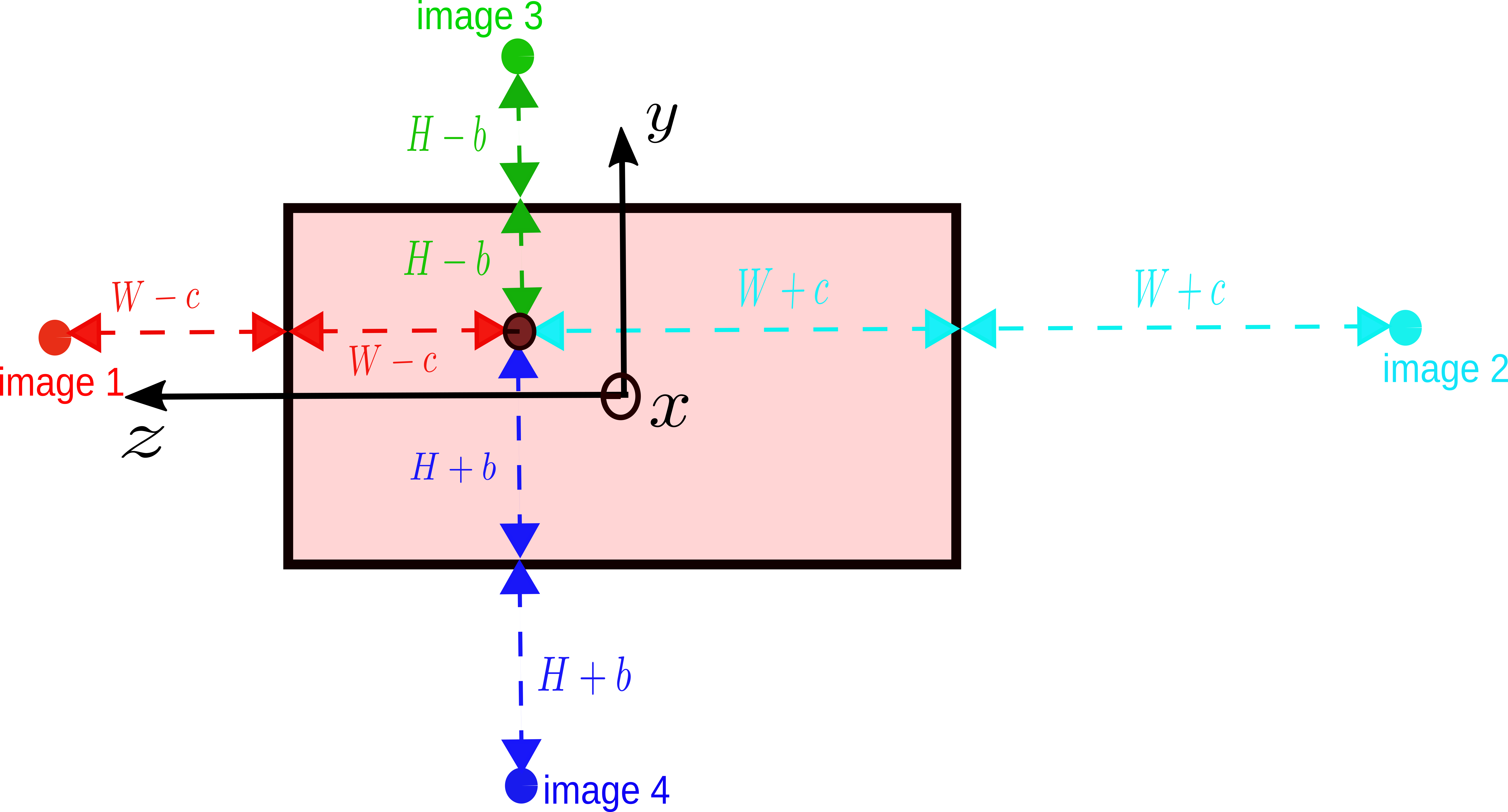}}
    \oversubcaption{0.03, 0.9}{}{fig:ffschem1}
    \oversubcaption{0.6, 0.9}{}{fig:ffschem2}
  \end{captivy}
\caption{(a) Schematic representation of the hydrodynamic image systems in a rectangular channel having a half height, $H$, and half width, $W$. A combination of force dipole of  strength $\alpha$ and source dipole of strength $q$, oriented along $\hat{\mathbf{e}}$, is placed at $(y,z) = (b,c)$. To the leading order, the hydrodynamic interaction of the microswimmer with each wall is approximated by a corresponding image system for this combination of force and source dipoles. The positions of the four image systems corresponding to the four walls are shown in the figure. (b) The $y-z$ plane of the channel and the distances of the squirmer and the image systems from the confining walls of the channel are shown here.}
    \label{fig:farfield}
\end{figure*}

In the multi-pole description of a force-free and torque-free microswimmer, the flow field generated by the squirmer is that of a stresslet in the leading order \citep{farfieldaccuracy}. 
The stresslet corresponds to two point forces acting at a sufficiently small separation distance. The point forces and their separation vector are aligned in the same direction. 
The strength of the stresslet, $\alpha$, is related to the second squirming mode as \citep{source_dipole}, 
\begin{align}
\alpha=-B_{2}(4\pi \mu a^{2})
\label{eqn:ffalpha}
\end{align}
The sign of the stresslet strength determines whether the squirmer is a puller ($\alpha<0$) or a pusher ($\alpha>0$). And of course, $\alpha = 0$ corresponds to a neutral swimmer.
It is now known that approximating the squirmer solely by a force dipole, or a stresslet, is insufficient to address the wall-confined dynamics for a finite-sized microswimmer especially in a confinement \citep{mathijssen2016upstream, de2019flow, dey2022oscillatory}. Hence, in order to consider the effects of the finite-size of the microswimmer inside a rectangular channel, we approximate the squirmer as a superposition of force and source dipoles. The strength of the source dipole $q$ is related to the first squirming mode as \citep{source_dipole}, 
\begin{align}
q=B_{1}(8\pi \mu a^{3}/3)
\label{eqn:ffq}
\end{align}

The aforementioned description of a squirmer works well when it is isolated, and located far from the confining boundaries of the channel. 
In the present work, the presence of a neighbouring channel wall is taken into account by considering the associated hydrodynamic interactions between the microswimmer and the wall using the method of images \citep{methodofimage, farfieldaccuracy}. 
We consider a squirmer, represented by the combination of stresslet of strength $\alpha$ and source dipole of strength $q$, located at $(y,z)=(b,c)$, and oriented along $\hat{\mathbf{e}}$. 
Fig.~\ref{fig:farfield} shows the schematic representing the $y-z$ cross-section of the channel, the location of the squirmer under consideration, and the corresponding positions of the total image systems (i.e. the image systems for stresslet and source dipole singularities) for each of the four confining walls of the rectangular channel. 

We first calculate the total velocity field due to the  image systems for the squirmer for each pair of walls, i.e. the left and right walls (image systems 1 and 2 in Fig.~\ref{fig:farfield}) and the top and bottom walls (image systems 3 and 4 in Fig.~\ref{fig:farfield}), following the methodology outlined in \citep{farfieldaccuracy, farfield}. 
Next, we use Fax\'en's laws \citep{kim2005microhydrodynamics} to calculate the wall-induced translational and angular velocities of the microswimmer corresponding to the velocity fields generated by the image systems for each pair of walls.
Applying Fax\'en's laws, the induced translational velocity, $\bar{\boldsymbol{U}}_{left/right}$, and angular velocity, $\bar{\boldsymbol{\Omega}}_{left/right}$, of the microswimmer due to the presence of the left and right walls can be calculated as \citep{farfieldaccuracy, farfield}
\begin{equation}
\begin{split}
\bar{\boldsymbol{U}}_{right/left}=&
\left\{ 
\bar{\alpha}
\left[
\mp \frac{3(1-3\sin^{2}(\theta_{z})}{8(\bar{W} \pm \bar{c})^2}
\pm \frac{(7-11\sin^{2}(\theta_{z})}{64(\bar{W} \pm \bar{c})^4}
\right]
\right.
\\
&
\left.
+
\bar{q}
\left[
\mp \frac{1}{(\bar{W}\pm \bar{c})^3}
 \pm \frac{1}{8(\bar{W} \pm \bar{c})^5}
\right]
\sin(\theta_{z})
\right\}
\hat{\mathbf{z}}
\\
+
&
\left \{
\bar{\alpha}
\left [ \pm
\frac{3}{8(\bar{W}\pm \bar{c})^2} \pm \frac{0.078125}{(\bar{W}\pm \bar{c})^4} 
\right]
\sin2(\theta_{z})\right.
\\
&\left.
+
\bar{q}
\left [ \mp
\frac{1}{4(\bar{W}\pm \bar{c})^3} \pm \frac{3}{16(\bar{W}\pm \bar{c})^5} 
\right]
\cos(\theta_{z})
\right \}\\
&
\times 
\left [
\frac{(\hat{\mathbf{e}}\cdot(\hat{\mathbf{x}}\hat{\mathbf{x}}+\hat{\mathbf{y}}\hat{\mathbf{y}}))}{|\hat{\mathbf{e}}\cdot(\hat{\mathbf{x}}\hat{\mathbf{x}}+\hat{\mathbf{y}}\hat{\mathbf{y}})|}
\right ]
\label{eqn:ffvelocitysingle}
\end{split}
\end{equation}
\begin{equation}
\begin{split}
\bar{\boldsymbol{\Omega}}_{right/left}=&
\left[ \mp
\frac{3\bar{\alpha}}{16} \frac{\sin2(\theta_{z})}{(\bar{W}\pm \bar{c})^{3}}
\pm
\frac{3\bar{q}}{8} \frac{\cos(\theta_{z})}{(\bar{W}\pm \bar{c})^{4}}
\right]
\\
&
\times 
\left[\frac{\hat{\mathbf{e}}\cdot(\hat{\mathbf{x}}\hat{\mathbf{x}}+\hat{\mathbf{y}}\hat{\mathbf{y}})}{|\hat{\mathbf{e}}\cdot(\hat{\mathbf{x}}\hat{\mathbf{x}}+\hat{\mathbf{y}}\hat{\mathbf{y}})|} \times (-\hat{\mathbf{z}})\right]
\label{eqn:ffangularsingle}
\end{split}
\end{equation}

\noindent
where the $\bar{\boldsymbol{U}}$ and $\bar{\boldsymbol{\Omega}}$ represent the dimensionless translational and angular velocities.
Here, each of the terms with $\pm/\mp$ represent two terms due to the image systems at the right and left walls. 
Specifically, the first and second signs represent the contributions from the right and left walls respectively. 
Strengths of force and source dipoles are non-dimensionalized using $a$ as the length scale and $U_f$ as the velocity scale to get $\bar{\alpha}$ and $\bar{q}$ respectively. 
$\theta_{z}$ is defined as $\theta_{z}=\arccos(\hat{\mathbf{e}}\cdot \hat{\mathbf{z}})$. 
Note that Eq. \ref{eqn:ffvelocitysingle} includes not only the lowest order terms obtained after applying Fax\'en's first law to the velocity field for the image systems, but also the higher order terms due to the Laplacian operator, i.e. terms proportional to $\frac{1}{(\bar{W}\pm \bar{c})^{4}}$ for the force-dipole image system and terms proportional to $\frac{1}{(\bar{W}\pm \bar{c})^{5}}$ for the source dipole image system. 

The translational and rotational velocities of the microswimmer induced by the top and bottom walls, $\bar{\boldsymbol{U}}_{top/bottom}$, can be also calculated using equations similar to Eq. \ref{eqn:ffvelocitysingle} and Eq. \ref{eqn:ffangularsingle}. 
Specifically, these can be obtained by changing $\hat{\mathbf{y}} \rightarrow \hat{\mathbf{z}}$, $\hat{\mathbf{z}} \rightarrow \hat{\mathbf{y}}$, $\bar{W} \rightarrow \bar{H}$, $\bar{c} \rightarrow \bar{b}$ and $\theta_{z} \rightarrow \theta_{y}$.
Finally, the total translational velocity $\bar{\boldsymbol{U}}_{channel}$ and angular velocity $\bar{\boldsymbol{\Omega}}_{channel}$ of the microswimmer due to the rectangular confinement can be calculated by adding the induced translational and angular velocities due to each wall (image system) as follows:
%

%\begin{widetext}
\begin{equation}
\begin{split}
\bar{\boldsymbol{U}}_{channel} =&   \hat{\mathbf{e}}\cdot(\hat{\mathbf{x}}\hat{\mathbf{x}}+\hat{\mathbf{y}}\hat{\mathbf{y}}+\hat{\mathbf{z}}\hat{\mathbf{z}}) +\sum_{all~walls}\bar{\boldsymbol{U}}_{wall}
%+\bar{\boldsymbol{U}}_{right~wall}+\bar{\boldsymbol{U}}_{top~wall}+\bar{\boldsymbol{U}}_{bottom~wall}
%\quad
%\quad
\\
\bar{\boldsymbol{\Omega}}_{channel}=& \sum_{all~walls} \bar{\boldsymbol{\Omega}}_{wall}
\label{eqn:ffvelocity}\\ 
\end{split}
\end{equation}
%\end{widetext}
where the first term in the expression of $\bar{\boldsymbol{U}}_{channel} $ is the self-propulsion velocity of the squirmer. 
It must be noted that the above formulation neglects the hydrodynamic interactions between the image systems, and the consequent effect on the dynamics of the microswimmer. 
Ideally, an infinite series of image systems needs to be considered in order to satisfy the boundary conditions at all the walls simultaneously.  
Therefore, Eq.~\ref{eqn:ffvelocity} gives only approximate expressions for calculating the dynamics of a finite-sized microswimmer in the rectangular channel.
The efficacy of this methodology cannot be commented upon \textit{a priori}, and is ascertained here by comparison with a full-scale numerical simulation. 
%%%%%%%%%%%%%%%%%%%%%%%%%%%%%%%%%%%%%%%%%%%%%%%%%%%%%%%%%%%%%%%%%%%%%%%%%%%%%%%%%%%%%%%%%%%%%%%%%%%%%%%%%%%%%%%%%%%%%
\subsection{Method 2: Numerical method}
 To obtain a rigorous description of the microswimmer dynamics in the rectangular channel, we use lattice-Boltzmann based numerical method and solve Eq.~\ref{eq:incompressibility} - \ref{eq:NS}.
%%%%%%%%%%%%%%%%%%%%%%%%%%%%%%%%%%%%%%%%%%%%%%%%%%%%%%%%%%%%%%%%%%%%%%%%%%%%%%%%%%%%%%%%%%%%%%%%%%%%%%%%%%%%%%%%%%%%%
% \subsubsection{Lattice-Boltzmann method (LBM)}
We follow the method discussed in \citet{ahana_activepassive} to calculate the spatio-temporal evolution of the fluid flow driven by the squirmer. The fluid motion and the dynamics of the micro-swimmer are coupled to each other as follows.
If the micro-swimmer has a translational velocity $\hat{\boldsymbol{U}}$, a rotational velocity $\hat{\boldsymbol{\Omega}}$ and a surface slip velocity $\hat{\boldsymbol{U}}_s$ (given by Eq.~\ref{eq:slip velocity}) then the velocity on the surface of the squirmer is given by,  
\begin{align}
\hat{\boldsymbol{u}}^s(\hat{\boldsymbol{r}}) = \hat{\boldsymbol{U}}_s (\hat{\boldsymbol{r}}) +  \hat{\boldsymbol{U}}+\hat{\boldsymbol{\Omega}}\times \hat{\boldsymbol{r}}.
\label{eqn:surface_velocity}
\end{align}
The boundary condition given by Eq.~\ref{eqn:surface_velocity} is incorporated in the lattice-Boltzmann frame work by implementing a mid grid bounce back scheme \citep{timm}.
At the same time, the dynamics of the micro-swimmer is  dictated by the stresses exerted by the fluid on the microswimmer.
The total force $\hat{\boldsymbol{F}}$ and torque $\hat{\boldsymbol{L}}$ on the micro-swimmer can be computed \citep{ahana2019confinement} and are used to update the translational and angular velocity of the micro-swimmer respectively, as described in \citet{nguyen2002lubrication},
\begin{align}
    m_s \hat{\boldsymbol{U}}(t+\Delta t) &=m_s \hat{\boldsymbol{U}}(t)+\Delta t \hat{\boldsymbol{F}}(t) 
    \label{eqn:velocity}\\
    I_s \hat{\boldsymbol{\Omega}}(t+\Delta t)&=I_s \hat{\boldsymbol{\Omega}}(t)+\Delta t \hat{\boldsymbol{L}}(t) 
    \label{eqn:omega}
\end{align}
where $m_s=(4/3)\pi \rho_s a^3 $ is the mass  and $I_s=(2/5) m_s a^2 $ is the moment of inertia of the squirmer.

%%%%%%%%%%%%%%%%%%%%%%%%%%%%%%%%%%%%%%%%%%%%%%%%%%%%%%%%%%%%%%%%%%%%%%%%%%%%%%%%%%%%%%%%%%%%%%%%%%%%%%%%%%%%%%%%%%%%%%%%%%%%%%%%%%%%%%%%%%%%%%%%%%%%%%%%%%%%%%%%%%%%%%%%%%%%%%%%%%%%%%%%%%%%%%%%%%%%%%%%%%%%%%%%%%%%%%%%%%%%%%%%%%%%%%%%%
\section{Results and discussion}
\label{sec:results}
Our results, presented in this section, are organized as follows. We start by computing the instantaneous velocity of the squirmer located at various positions inside the channel while its orientation is parallel to the channel axis. These results will enable us to understand the origin of the two-dimensional trajectories that squirmers follow inside the channel, which are discussed subsequently. We will then examine the  three-dimensional trajectories of squirmers in square channels ($AR=0$), and finally in rectangular channels ($AR>0$). The principle of superposition is described afterwards to understand the apparently complex 3D trajectories.
%%%%%%%%%%%%%%%%%%%%%%%%%%%%%%%%%%%%%%%%%%%%%%%%%%%%%%%%%%%%%%%%%%%%%%%%%%%%%%%%%%%%%%%%%%%%%%%%%%%%%%%%%%%%%%%%%%%%%
\subsection{Instantaneous velocity of squirmers}
\begin{figure*}
    \centering
  \includegraphics[clip,trim=0cm 0cm 0cm 0cm,width=.31\textwidth]{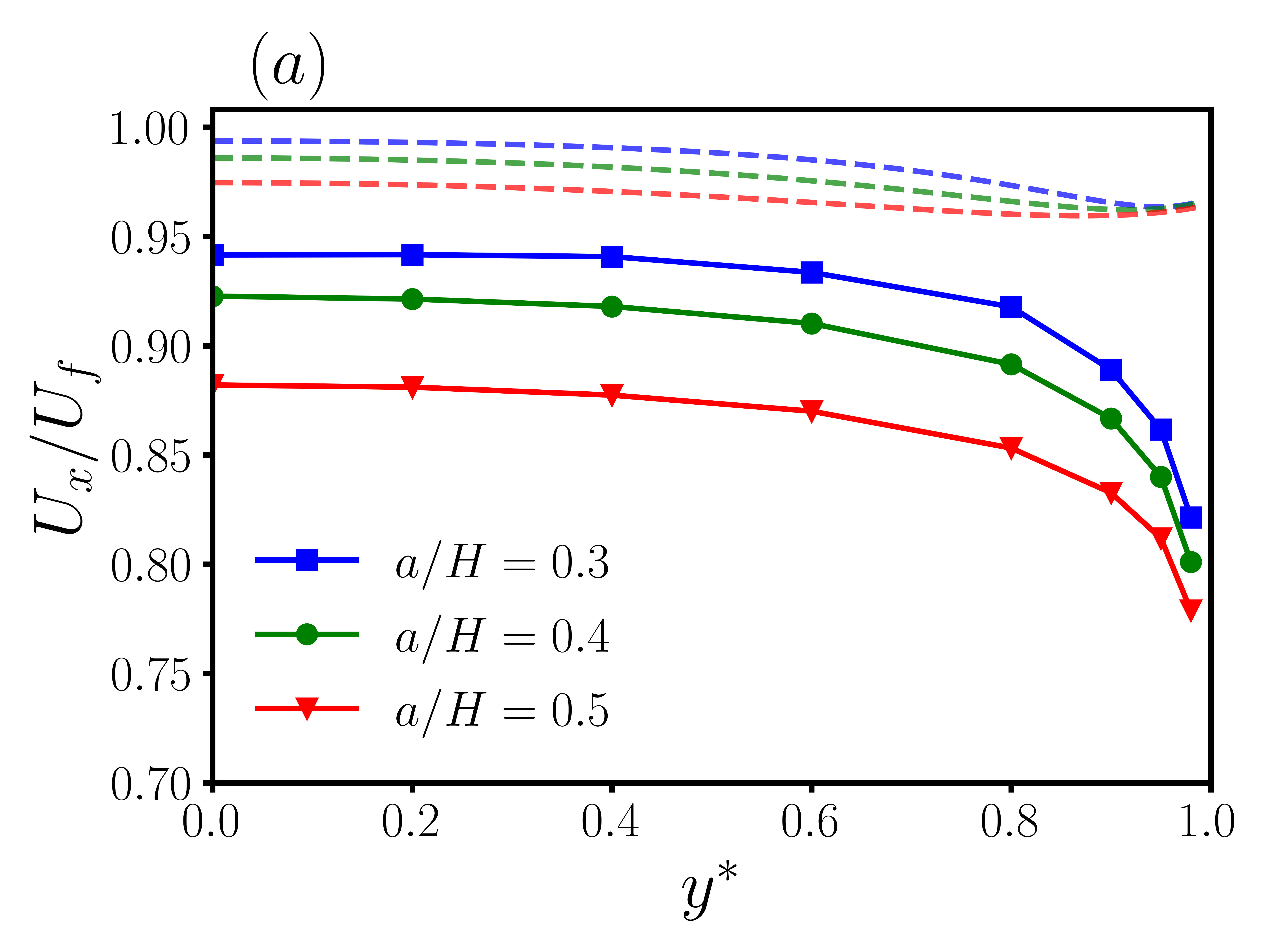}
  %\label{fig:instantaneous_velocity1}
  \hspace{.25cm}
 \includegraphics[clip,trim=0cm 0cm 0cm 0cm,width=.31\textwidth]{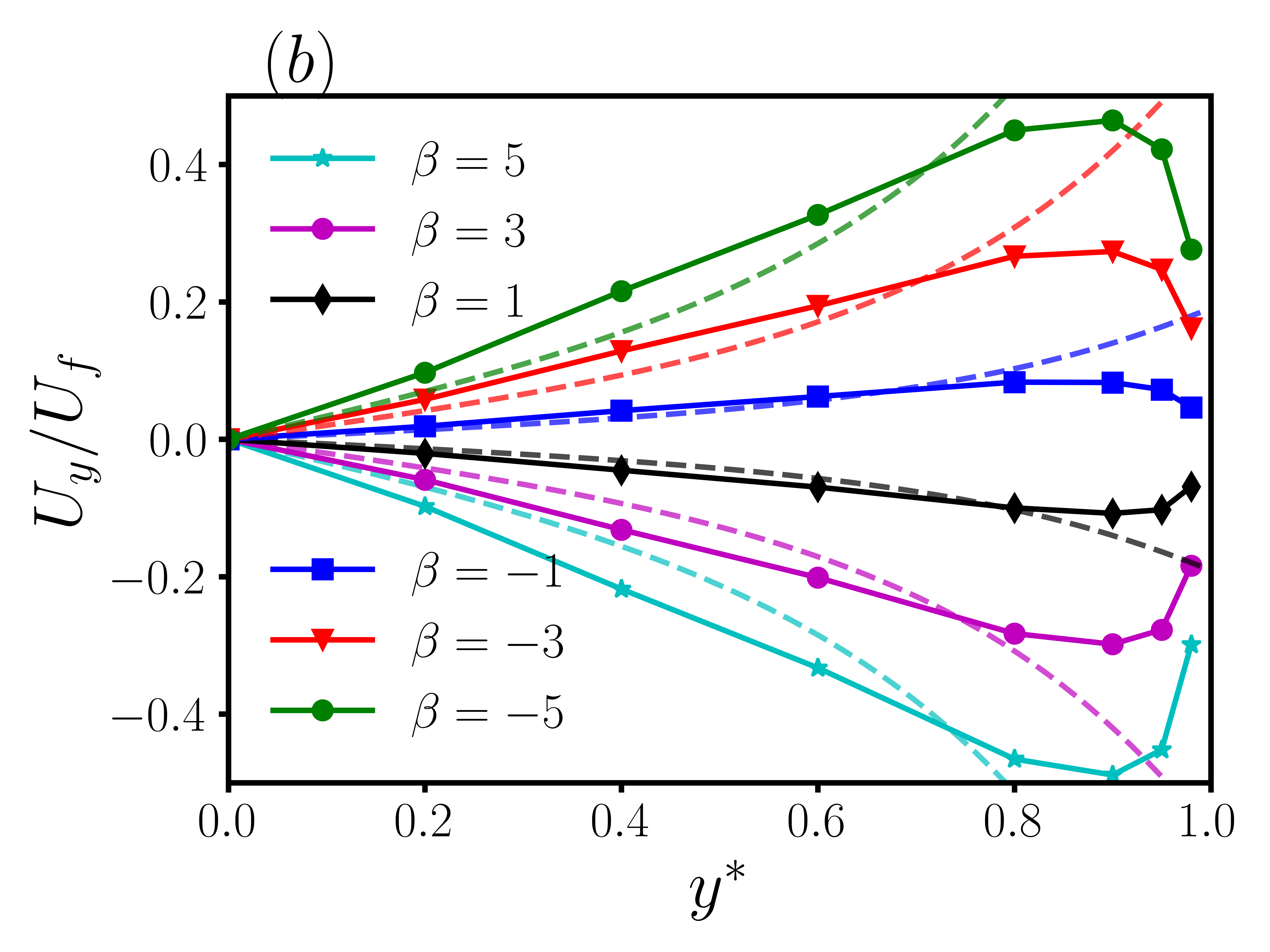}
 %\label{fig:instantaneous_velocity2}
 \hspace{.25cm}
 \includegraphics[clip,trim=0cm 0cm 0cm 0cm,width=.31\textwidth]{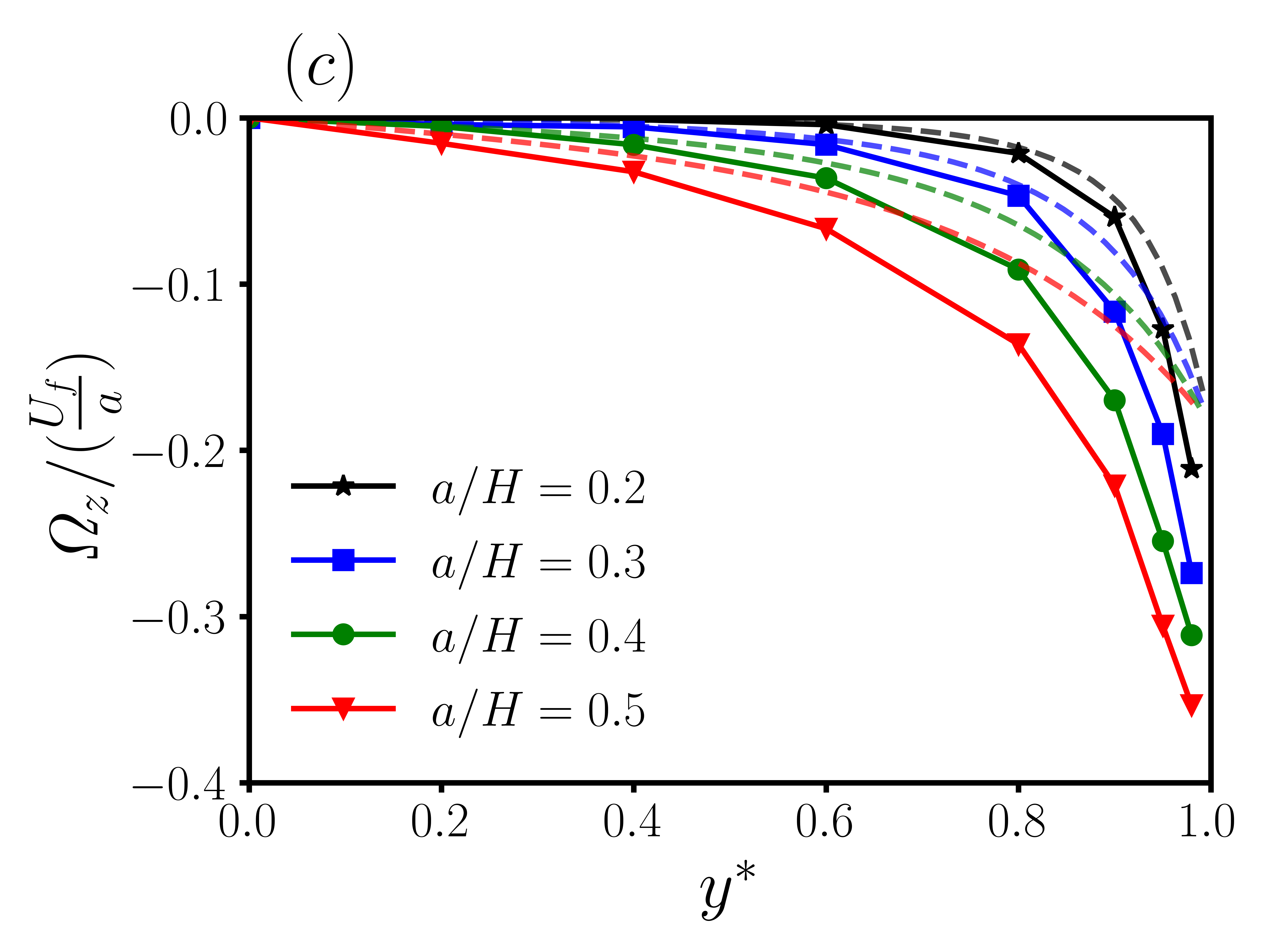}
\caption{Instantaneous swimming velocity of a squirmer in a square channel when it is initially located at $(0, y^{*}, 0)$ with its orientation parallel to the channel axis ($\theta=0$). (a) Swimming velocity in the $x$-direction, $U_{x}$, scaled by the swimming speed in unconfined fluid $U_{f}$, for various values of $a/H$. (b) Swimming speed $U_{y}$ in the $y$-direction, scaled by $U_{f}$, for different values of $\beta$ when $a/H = 0.3$. (c) The instantaneous angular velocity of the swimmer in the direction normal to the plane of locomotion $\Omega_{z}$, scaled by the quantity $U_{f}/a$, for different values of $a/H$. Plot (c) is same for all types of squirmers, since the induced  angular velocity due to $B_2$ mode will vanish for $\theta =0$. Dashed lines here (and throughout this study) represent the results from the far-field expressions. The far-field approximation provides qualitatively accurate predictions for $U_x$, $U_y$ and $\Omega_z$.}
    \label{fig:instantaneous_velocity}% of $U_x$ and $\Omega_z$. $U_y$ calculated from far-field expressions and numerical simulations match well up to $y^{*} = 0.5$.
\end{figure*}

The instantaneous swimming velocity of a squirmer in the square channel, oriented parallel to the channel centerline ($\theta=0$) is calculated as a function of distance from the centerline $y^{*}$ and for different confinement ratios $a/H$. The results are plotted as follows - the instantaneous velocity parallel to the channel axis $U_{x}$ and perpendicular to the channel axis $U_{y}$ are plotted in Fig.~\ref{fig:instantaneous_velocity}(a) and \ref{fig:instantaneous_velocity}(b) respectively, and in Fig.~\ref{fig:instantaneous_velocity}(c) the angular velocity in $z$-direction ($\Omega_{z}$) is depicted. 

The swimming velocity parallel to the channel axis, $U_{x}$, is independent of the second mode of squirming $B_{2}$. 
Hence, Fig.~\ref{fig:instantaneous_velocity}(a) is same for pushers, pullers and neutral swimmers. 
This happens due to the fore-aft symmetry of the second squirming mode, $B_{2}\sin (2\theta)$, which does not generate any wall-induced velocity along the channel length for a microswimmer oriented parallel to the channel axis \citep{Berk}. 
As seen in Fig.~\ref{fig:instantaneous_velocity}(a), $U_{x}$ decreases with increase in confinement, $a/H$, and with increasing distance from the centerline, $y^{*}$. 
The sharp decrease in $U_{x}$ for $y^{*} > 0.8$ is an indication of strong interaction of the squirmer with the nearby wall. 
The far-field approximation predicts qualitatively similar behaviour for $y^{*} < 0.9$, which is shown as dashed lines in Fig.~\ref{fig:instantaneous_velocity}(a). 
Far-field approximation fails for $y^{*}>0.9$ since these approximations are highly unstable when the microswimmer comes in the immediate vicinity of a wall.
Unlike $U_x$, the instantaneous velocity in $y$-direction $U_y$ is dependent on $\beta$.
Therefore, $U_y$ is plotted  against the squirmer location, $y^{*}$, by varying $\beta$ for a fixed confinement ratio of $a/H = 0.3$.  
The magnitude of $U_{y}$ is equal, but the signs are opposite for pushers and pullers of same strength.
Pullers ($\beta > 0$) have a negative velocity along $y$-direction ($U_{y}< 0$), and accordingly, they are repelled from the nearest wall. 
In contrast, pushers ($\beta < 0$) are attracted towards the nearest boundary wall ($U_{y} > 0$).
Similar results are also obtained from the far-field calculations (Eq.~\ref{eqn:ffvelocity}) for for $y^{*} < 0.8$.
As the squirmer gets very close to the wall ($y^{*}>0.8$), the resulting hydrodynamic interaction reduces the strength of $U_{y}$ substantially as shown in Fig.~\ref{fig:instantaneous_velocity}(b). 
It may also be seen that $|U_{y}|$ increases as the stresslet strength of the pushers and pullers increases. 

As shown in Fig.~\ref{fig:instantaneous_velocity}(c) hydrodynamic interaction of the squirmer with the channel boundaries also results in a wall induced angular velocity which changes the orientation of squirmers. 
The angular velocity $\Omega_{z}$ increases with increase in both $y^{*}$ and $a/H$, a manifestation of strong hydrodynamic interaction between the nearest wall and the finite-sized squirmer. 
Positive (negative) value of angular velocity corresponds to anticlockwise (clockwise) rotation of the squirmer. 
Far-field expressions predict similar variations in $\Omega_z $ with increase in $y^{*}$ and confinement. 
The results indicate that all squirmers rotate away from the closest wall. 
This is a consequence of the hydrodynamic interaction due to the finite-size of the microswimmer, as captured by the first squirming mode or the source dipole in the far-field calculation. 
The angular velocity due to the source dipole has a $\cos \theta$ dependence, which generates a non-zero angular velocity even when the squirmer is oriented parallel to the channel axis. 
%%%%%%%%%%%%%%%%%%%%%%%%%%%%%%%%%%%%%%%%%%%%%%%%%%%%%%%%%%%%%%%%%%%%%%%%%%%%%%%%%%%%%%%%%%%%%%%%%%%%%%%%%%%%%%%%%%%%%
\subsection{2D motion in square channels}
\label{subsec:2d_results}
In this section, we focus on the two-dimensional trajectories of the squirmers inside a square channel. The initial location and orientation of the squirmer are identified by specifying a subscript $i$ to the relevant variable. In order to obtain two-dimensional trajectories, the squirmer is initially placed in the $x$-$y$ plane with its orientation vector lying in this plane, \textit{i.e.,} $z_i^{*}=0$, $\delta_{i}=0$. The symmetry of the system ensures that the squirmer remains in the $x$-$y$ plane as it propels and generates a two-dimensional trajectory. 
%%%%%%%%%%%%%%%%%%%%%%%%%%%%%%%%%%%%%%%%%%%%%%%%%%%%%%%%%%%%%%%%%%%%%%%%%%%%%%%%%%%%%%%%%%%%%%%%%%%%%%%%%%%%%%%%%%%%%
\subsubsection{Oscillatory trajectories of pushers}
\label{subsubsec:oscillation_pushers}
\begin{figure}[!tp]
    \centering
 \includegraphics[clip,trim=0cm 0cm 0cm 0cm,width=.5\textwidth]{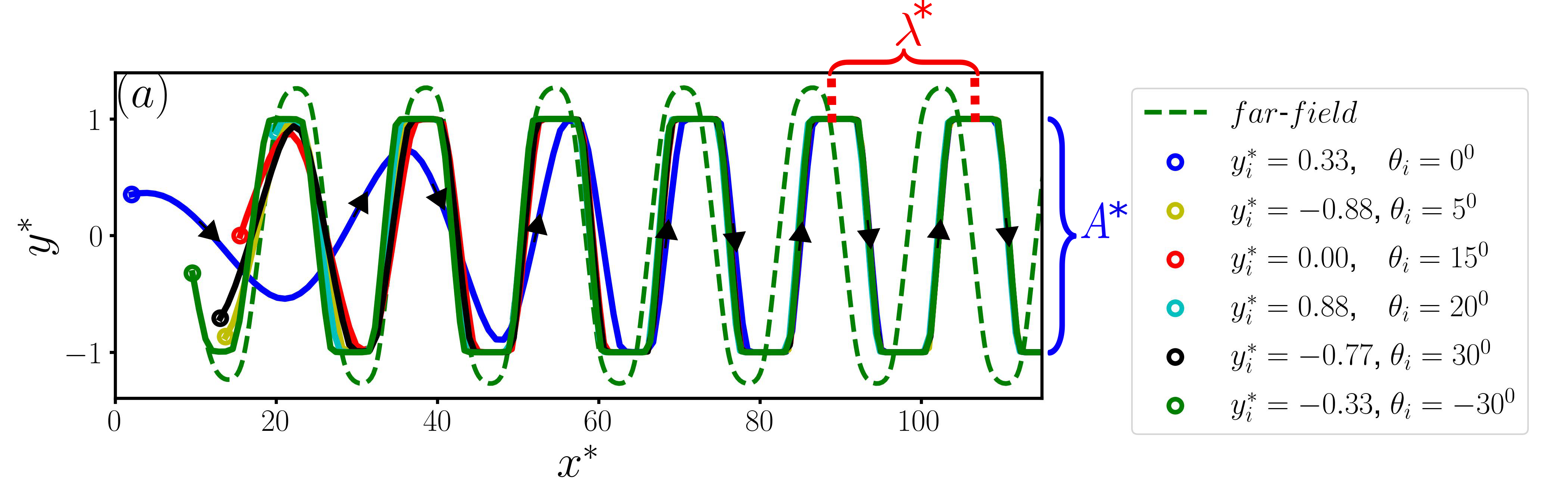}
  \includegraphics[clip,trim=0cm 0cm 0cm 0cm,width=.235\textwidth]{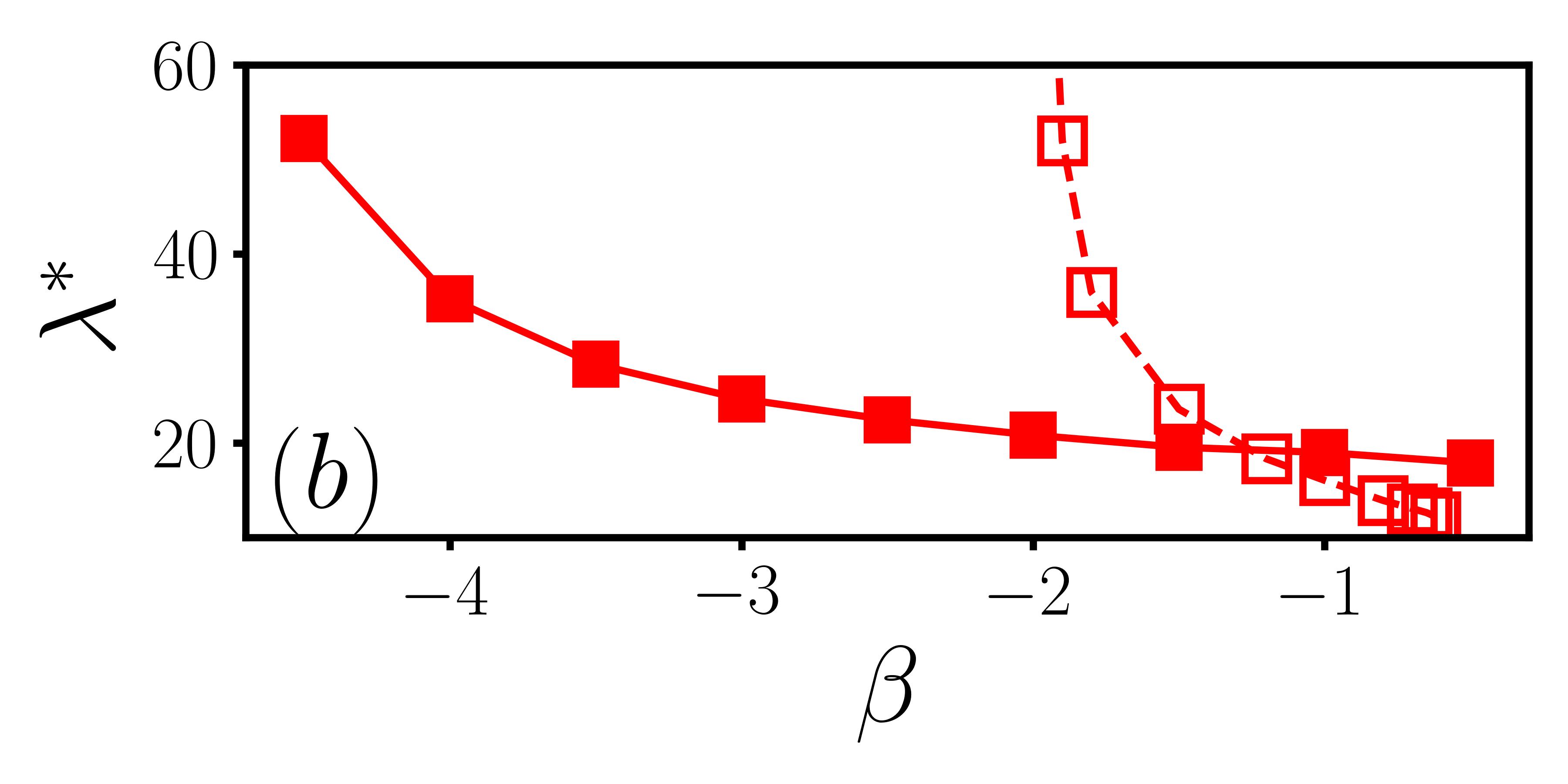}
\includegraphics[clip,trim=0cm 0cm 0cm 0cm,width=.235\textwidth]{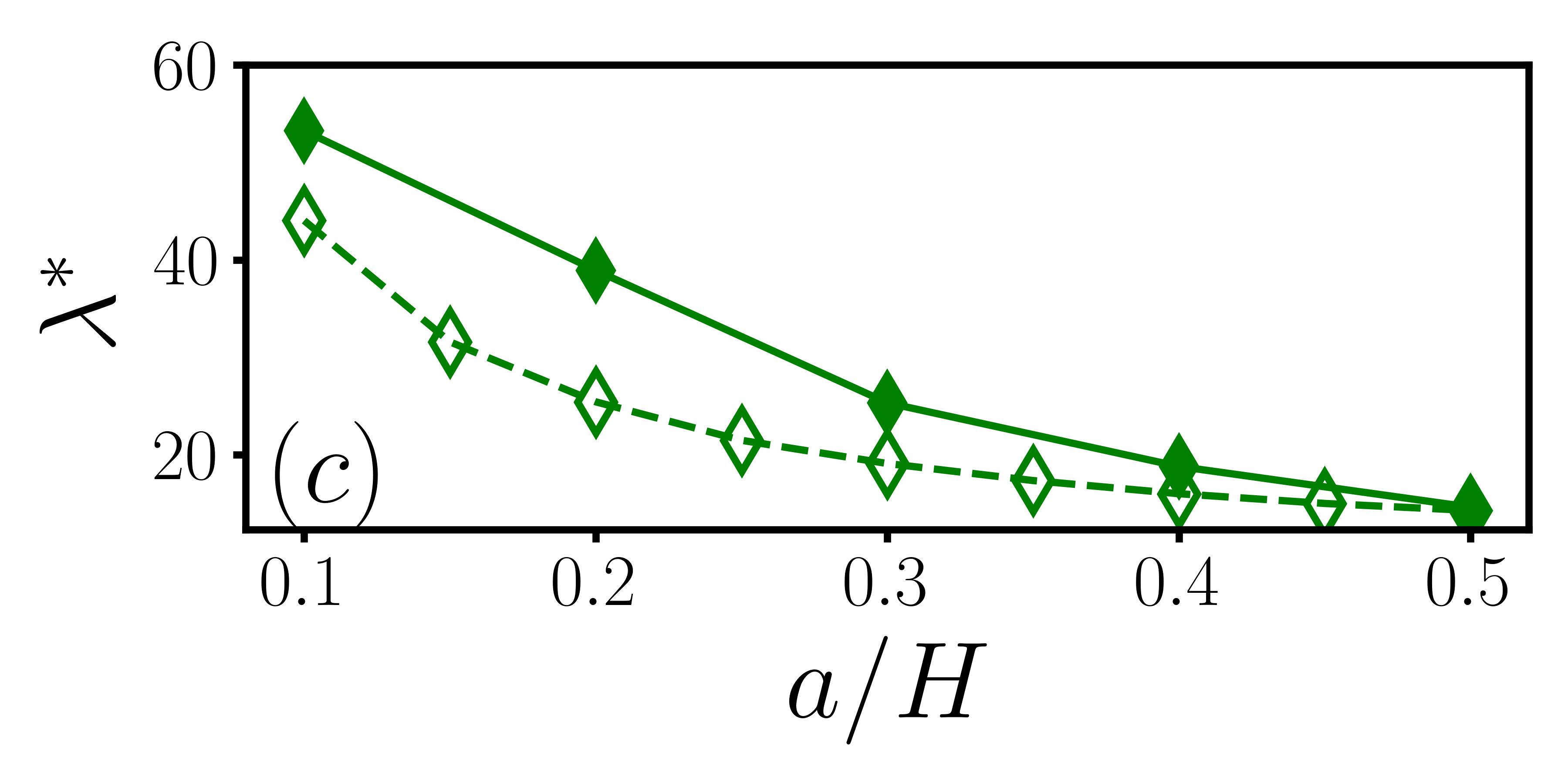}
%}
\caption{(a) Oscillatory motion of pushers in the $x$-$y$ plane. Simulations with different initial conditions ($y^{*}_{i},\theta_{i}$) give trajectories of same wavelength ($\lambda^{*}$) and amplitude ($A^{*}$). Variation of $\lambda^{*}$ with (b) the strength of the second mode of squirming, $\beta$ and (c) the confinement ratio $a/H$.} 
%Initial conditions  are same for both plots ($y^{*}_{i}=0$, $\theta_{i}=30^{0}$).}
    \label{fig:Pushers_oscillation}
\end{figure}
Our simulations show that pushers make periodic, oscillatory trajectories in the center plane of the channel as plotted in Fig.~\ref{fig:Pushers_oscillation}(a). 
As the pusher approaches the wall, angular velocity $\Omega_{z}$ induced by the hydrodynamic interaction with the wall (Fig.~\ref{fig:instantaneous_velocity}(c)) rotates the squirmer away from the nearest wall. 
This is the key reason for the existence of oscillatory trajectory of pushers.  
During such motion, pushers swim across the entire channel width, \textit{i.e} the channel width determines the amplitude, $A^{*}$, of the oscillatory trajectory. 
At the same time, the wavelength ($\lambda^{*}$) of the trajectory is set by the channel width, confinement, and $\beta$. 
Both $\lambda^{*}$ and $A^{*}$ of the trajectories are found to be independent of initial conditions. 
The oscillatory trajectories are also qualitatively similar to those generated by circular squirmers in two-dimensional channels \citep{ahana2019confinement}.

The prediction from the far-field calculation is plotted in Fig.~\ref{fig:Pushers_oscillation}(a) as a dashed line. 
The far-field analysis also gives an oscillatory trajectory, but with a slightly different amplitude and wavelength. 
The trajectories from the far-field calculations, for a given $\beta$ and $a/H$, are also independent of initial conditions. 
It is important to note here that we have to consider both the force and source dipoles in the far-field calculations to reproduce the oscillatory trajectory of a confined microswimmer, as predicted by the numerical simulation. 
Considering solely the force dipole singularity results in a sliding trajectory along the wall for a pusher, which will be an erroneous prediction for a microswimmer in a confinement.
Also, the higher-order terms, i.e. the terms $\sim 1/r^{4}$ and $\sim 1/r^{5}$ in Eq.~\ref{eqn:ffvelocity}, are essential to reproduce the oscillations of pushers. 
The contributions to the velocity from these higher-order terms are opposite to the first-order terms and become significant when the pusher approaches the wall. 
These effects, along with the finite angular velocity induced by the source dipole singularity, prevent the unphysical crossing of the rigid wall by the microswimmer.

On analyzing the trajectories further, we find that $\lambda^{*}$ exhibits a strong dependence on the dipole strength of pushers and the confinement ratio. 
This dependence, for both the results from the numerical simulations and the far-field calculations, is illustrated in Fig.~\ref{fig:Pushers_oscillation}(b) and (c). 
Strong pushers spend more time closer to the wall due to their stronger velocity component towards the wall, and make larger escape angles. 
Accordingly, the wavelength of the trajectory increases with increase in the strength of pushers (Fig.~\ref{fig:Pushers_oscillation}(b)). 
In contrast, $\lambda^{*}$ decreases with increase in confinement ratio, $a/H$ (Fig.~\ref{fig:Pushers_oscillation}(c)).
Strong confinement results in stronger hydrodynamic interaction of the finite-sized microswimmer with the walls, which manifests in greater angular velocity and stronger repulsion from the wall; accordingly, pushers escape quickly. 
%%%%%%%%%%%%%%%%%%%%%%%%%%%%%%%%%%%%%%%%%%%%%%%%%%%%%%%%%%%%%%%%%%%%%%%%%%%%%%%%%%%%%%%%%%%%%%%%%%%%%%%%%%%%%%%%%%%%%
\subsubsection{Periodic oscillation of neutral swimmers}
\begin{figure}[!tp]
    \centering
\includegraphics[clip,trim=0cm 0cm 0cm 0cm,width=.5\textwidth]{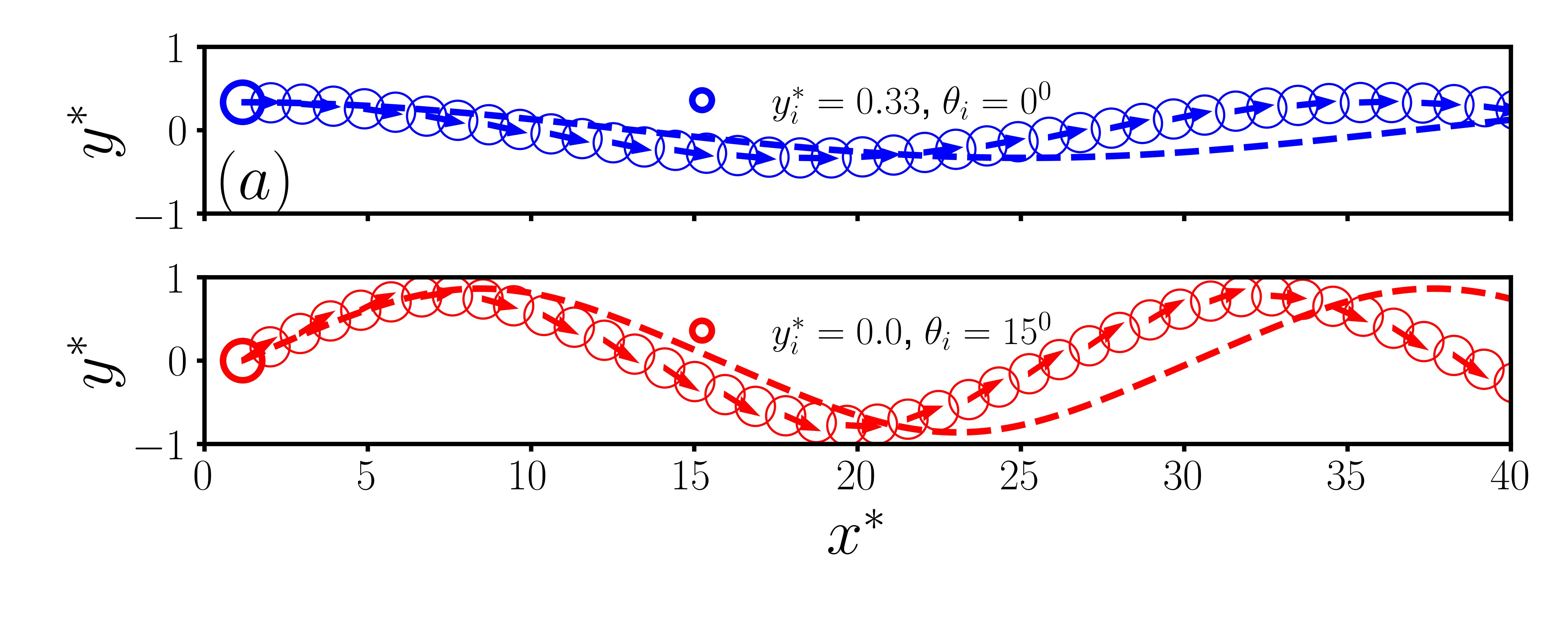}
  \hspace{.1cm}
         \includegraphics[clip,trim=0cm 0cm 0cm 0cm,width=.25\textwidth]{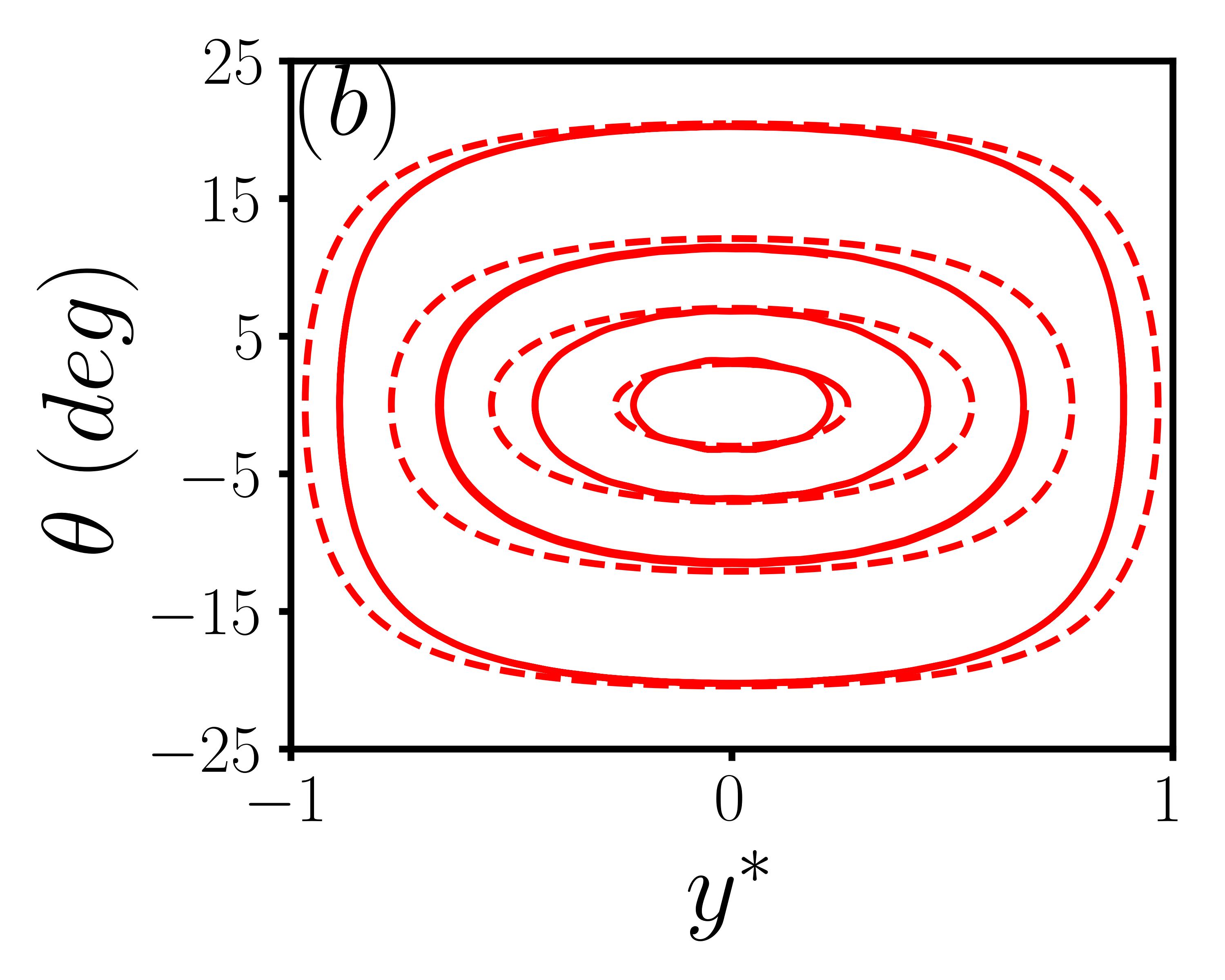}
\caption{Periodic oscillations of channel confined neutral swimmers: (a) Oscillatory trajectories in $x-y$ plane corresponding to two different initial conditions. The arrows show the instantaneous orientation of the squirmer. (b) Phase diagram in ($y^{*},\theta$) plane showing the periodic oscillation of neutral swimmers as stable limit cycles. Each closed curve represents oscillation corresponding to a particular initial position and orientation.}
    \label{fig:2d_neutral}
\end{figure}

Next, we move on to the trajectory analysis of a neutral swimmer. 
We find that a neutral swimmer always follows a periodic oscillatory trajectory in the channel with $\lambda^{*}$ and $A^{*}$ that are \textit{dependent} on initial conditions. 
Fig.~\ref{fig:2d_neutral}(a) illustrates an example, where trajectories of a neutral swimmer for two different initial conditions $y^{*}_{i}=0.33, \theta^{*}_{i}=0^{0}$ (top ) and  $y^{*}_{i}=0.0, \theta^{*}_{i}=15^{0}$ (bottom) are shown. 

Unlike pushers, neutral swimmers immediately escape as they approach the walls.
Again, this is a consequence of the non-zero angular velocity
$\Omega_{z}$ which changes the orientation of the squirmer as it approaches the wall. 
The amplitude of the trajectory is constant over time, and is two times the initial distance from the channel centerline ($A^{*}=2y^{*}_{i}$) if the squirmer is initially oriented parallel to the wall ($\theta_i=0$). These results are qualitatively consistent with the behavior of neutral swimmers inside a capillary tube \citep{capillary}, and circular neutral swimmers in 2-dimensional channels \citep{ahana2019confinement}. 
In this case also, the far-field predictions are in good agreement with the simulation data (Fig.~\ref{fig:2d_neutral}(a)). 
It is to be noted that the strength of the force dipole is zero for neutral swimmers. 
The far-field velocities have contributions only from the source dipole, and  are capable of reproducing oscillatory trajectories with wavelength and amplitude chosen by the initial conditions.

The oscillatory trajectories of neutral swimmers can be characterised by two parameters: the distance from the centerline ($y^{*}$) and the orientation of the squirmer ($\theta$). 
Therefore, the oscillatory trajectories corresponding to different initial conditions can be easily represented in a phase portrait in $(y^{*},\theta)$ plane as in Fig.~\ref{fig:2d_neutral}(b). 
For every initial condition except ($y^*_i=0$, $\theta_i=0$), the neutral swimmer exhibits a periodic oscillation which represents a stable limit cycle in the phase portrait. 
The initial condition ($y^*_i=0$, $\theta_i=0$) is an unstable fixed point that corresponds to the translation of the squirmer through the centerline of the channel without any deflection. 
However, a slight perturbation will produce an oscillatory trajectory. 
Even though both pushers and neutral swimmers follow periodic trajectories, as we see in the later sections, the three-dimensional motion of neutral swimmers are more challenging to understand due to the strong dependence of their trajectories on the initial conditions.
%
%
%%%%%%%%%%%%%%%%%%%%%%%%%%%%%%%%%%%%%%%%%%%%%%%%%%%%%%%%%%%%%%%%%%%%%%%%%%%%%%%%%%%%%%%%%%%%%%%%%%%%%%%%%%%%%%%%%%%%%
\subsubsection{Sliding motion of pullers}
\label{subsubsec:sliding_pullers} 
In contrast to pushers and neutral swimmers, pullers exhibit sliding motion in the channel. This sliding can be either through the channel centerline or along a straight path close to the walls depending on the strength of the puller ($\beta$) and the initial conditions.

\begin{figure}
    \centering
 \includegraphics[clip,trim=0cm 0cm 0cm 0cm,width=.238\textwidth]{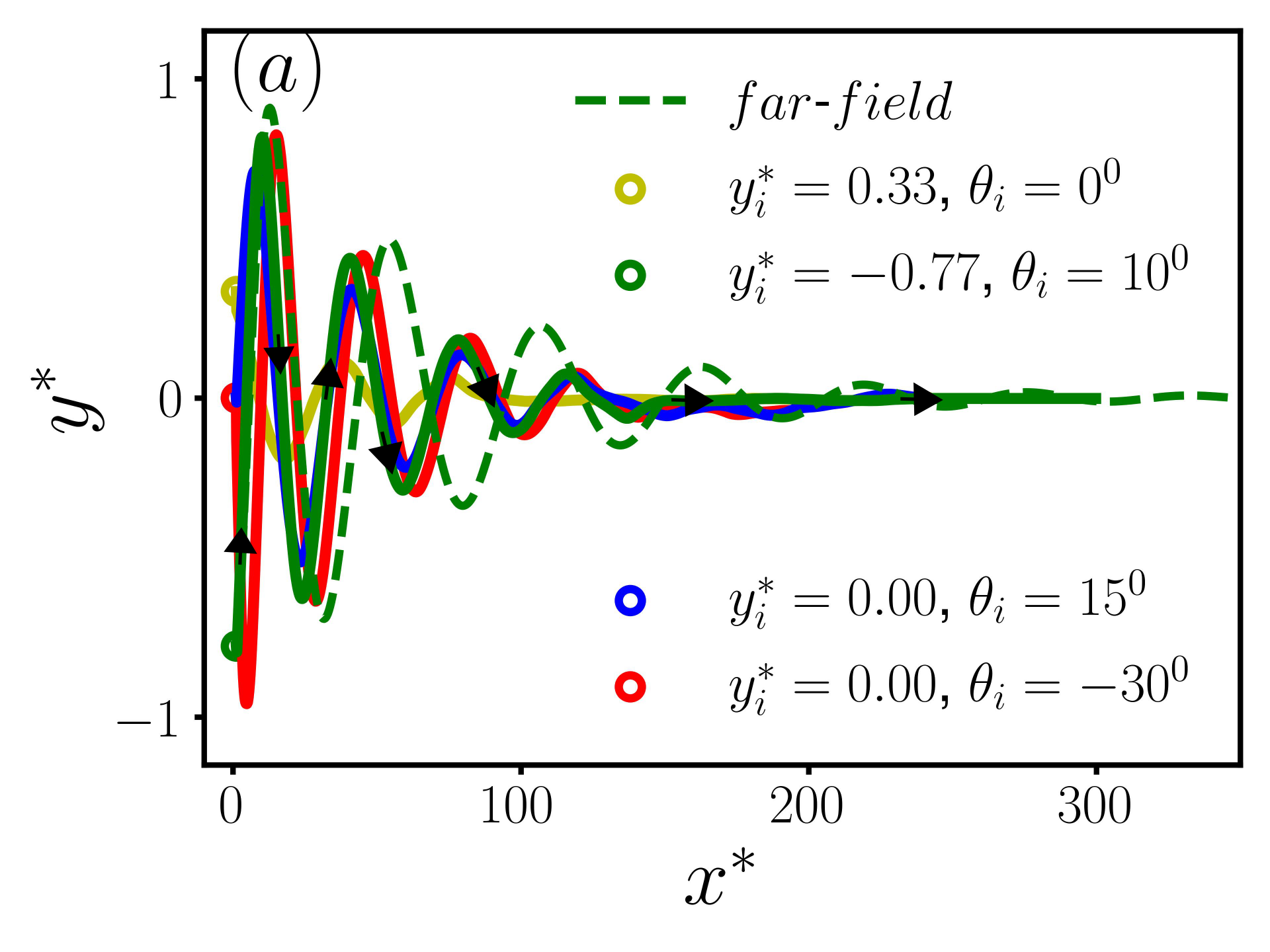}
%  \hspace{.1cm}
          \includegraphics[clip,trim=0cm 0cm 0cm 0cm,width=.238\textwidth]{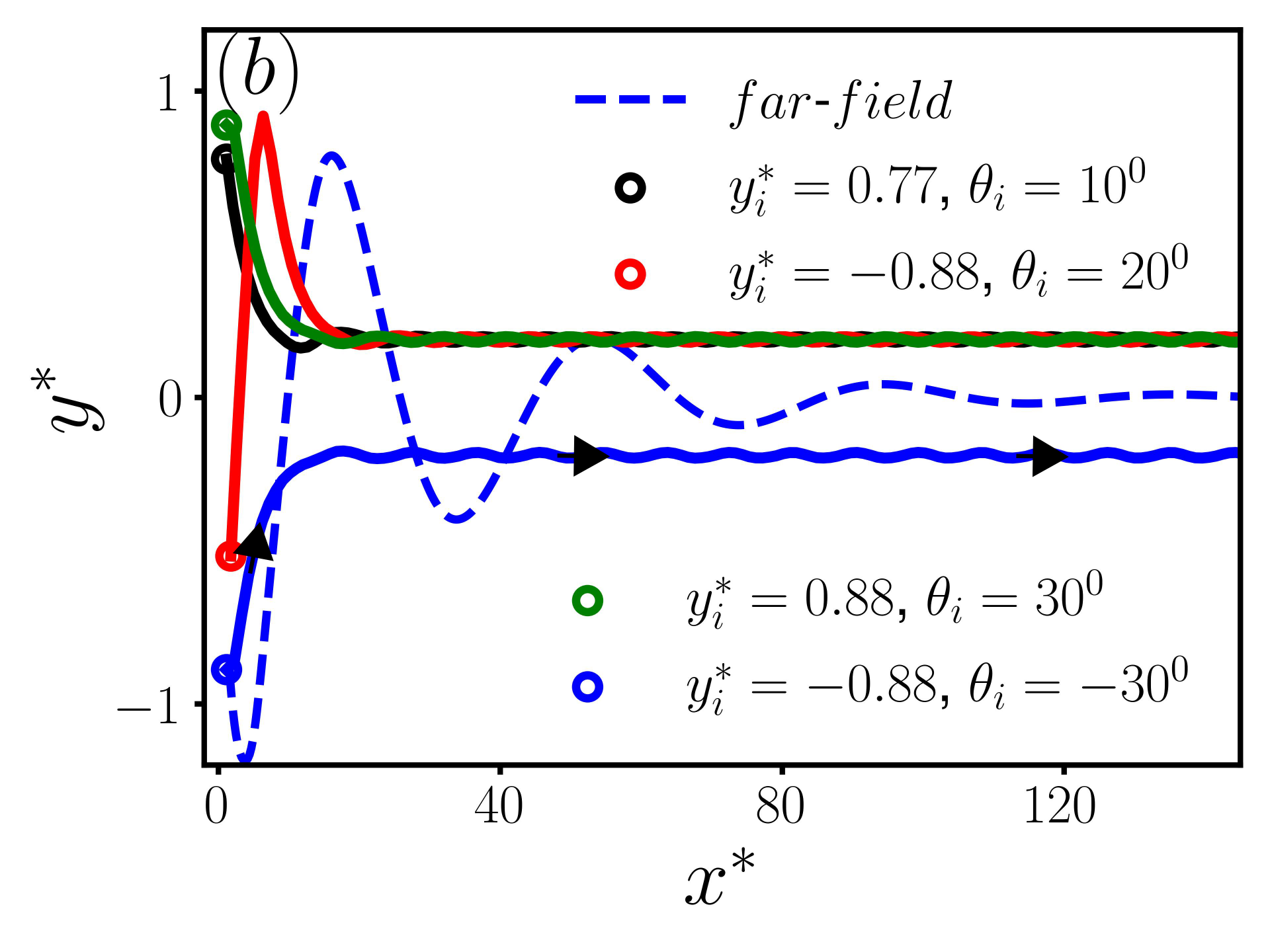}
\caption{Sliding motion of pullers in $x$-$y$ plane for various initial conditions for (a) $\beta = 1$ and (b) $\beta = 5$. The swimmers exhibit initial, transient oscillations and finally they slide along the channel centerline as in (a) or closer to one of the channel walls as in (b) depending on their strength.}
    \label{fig:pullers}
\end{figure}

Fig.~\ref{fig:pullers}(a) illustrates sliding of a puller ($\beta=1$) through the channel center-line. 
Before reaching this steady state, pullers exhibit transient oscillations, the amplitude of which gradually decreases, and finally they swim in a straight-line trajectory. 
During this straight line motion through the channel center-line, they maintain an orientation towards the +ve $x$-axis.
Interestingly, stronger pullers, say $\beta=5$ as shown in Fig.~\ref{fig:pullers}(b), after exhibiting damped oscillations end up in a path parallel to the channel axis, but not along the center-line. 
This trajectory is nearer to one of the walls depending on the initial conditions, and the squirmer will be oriented slightly towards the nearest wall. 
In this configuration both the wall induced translational velocity normal to the wall and the angular velocity are simultaneously zero, which sustains the straight line motion. 

These results imply the existance of a critical strength of pullers, $\beta_c \approx 3.55$, below which pullers slide through the channel center-line and above which they slide near to one of the walls. 
A similar observation was also made for microswimmers in a capillary \citep{capillary}. 
Our simulations show that a strong puller converges to its equilibrium trajectory close to a wall quickly (Fig.~\ref{fig:pullers}(b)), while a weak puller ($\beta<3$) takes a longer time to attain the sliding trajectory along the channel center-line (Fig.~\ref{fig:pullers}(a)). 
Note that the initial condition ($y^{*}_{i}=0,\theta_{i}=0$) is a special condition which results in the straight line translation of the puller through the channel centerline.

As in the previous sections, the far-field analysis successfully predicts the trajectory of weak pullers through the channel center-line. 
However, it fails to predict the sliding motion near one of the walls for a strong puller (Fig.~\ref{fig:pullers}(b)) illustrating the role of source dipole in determining the behaviour of a puller when it is close to the confining surface. 
As we increase the strength of the force dipole, we find a transition (not shown) from sliding through the center-line to a trajectory where pullers orient perpendicular to the nearest wall and get trapped by that wall (hovering behavior). 
%However, these calculations cannot capture the sliding motion near one of the walls.
%
%
%
%
%%%%%%%%%%%%%%%%%%%%%%%%%%%%%%%%%%%%%%%%%%%%%%%%%%%%%%%%%%%%%%%%%%%%%%%%%%%%%%%%%%%%%%%%%%%%%%%%%%%%%%%%%%%%%%%%%%%%%
\subsubsection{Phase diagram and the role of initial conditions}
\label{subsubsec:phase}
In this section, we construct a phase diagram showing the behavior of a squirmer for different strength, $\beta$, and confinement ratio, $ a/H$. We also report that the phase diagram depends on the initial conditions by comparing it for two different initial conditions: initial orientations $\theta_{i}=60^{\circ}$ and $\theta_{i}=30^{\circ}$ for fixed initial position $y^{*}_{i}=0$.

\begin{figure}
    \centering
 \includegraphics[clip,trim=0cm 0cm 0cm 0cm,width=.238\textwidth]{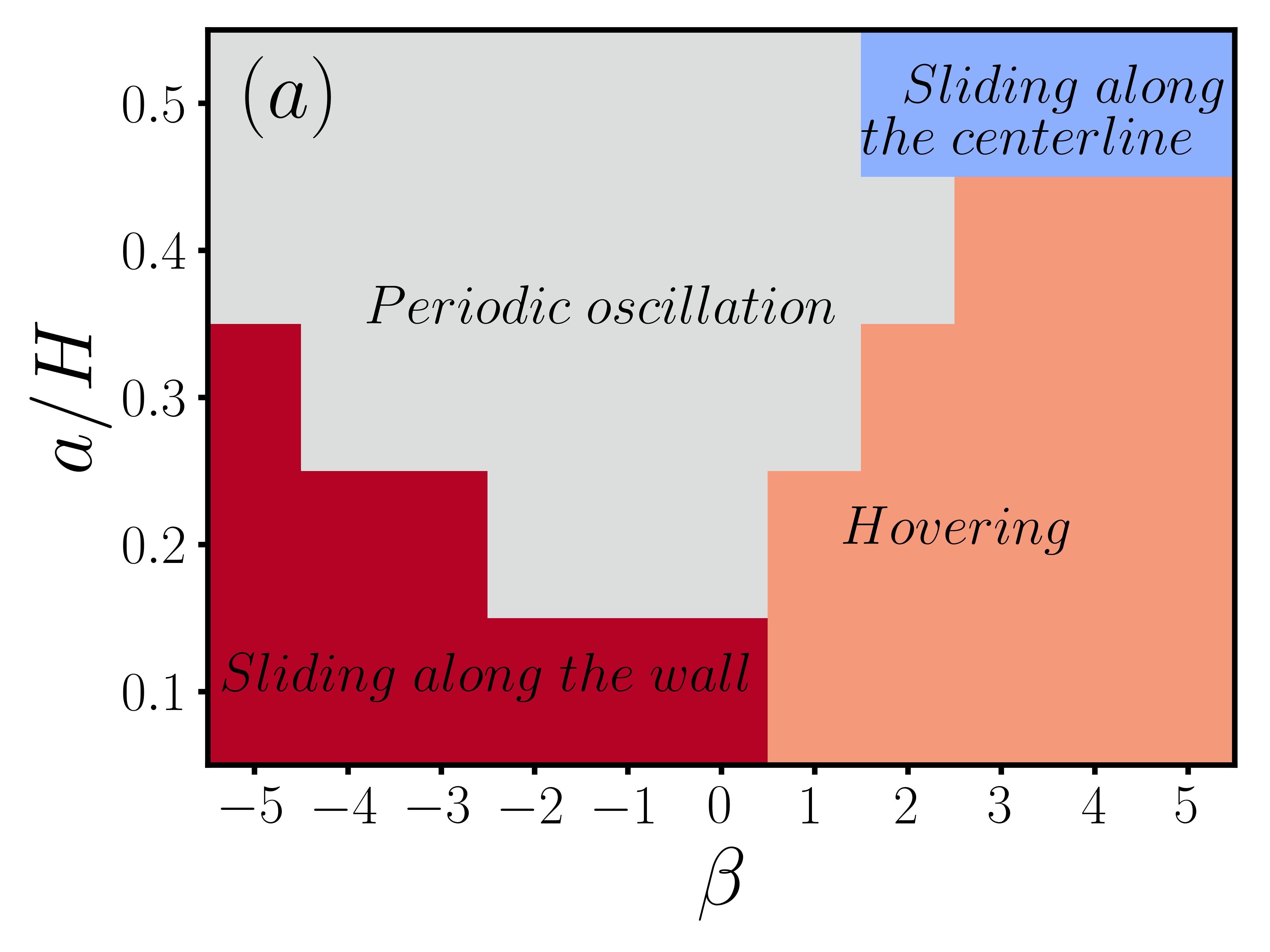}
%  \hspace{.1cm}
          \includegraphics[clip,trim=0cm 0cm 0cm 0cm,width=.238\textwidth]{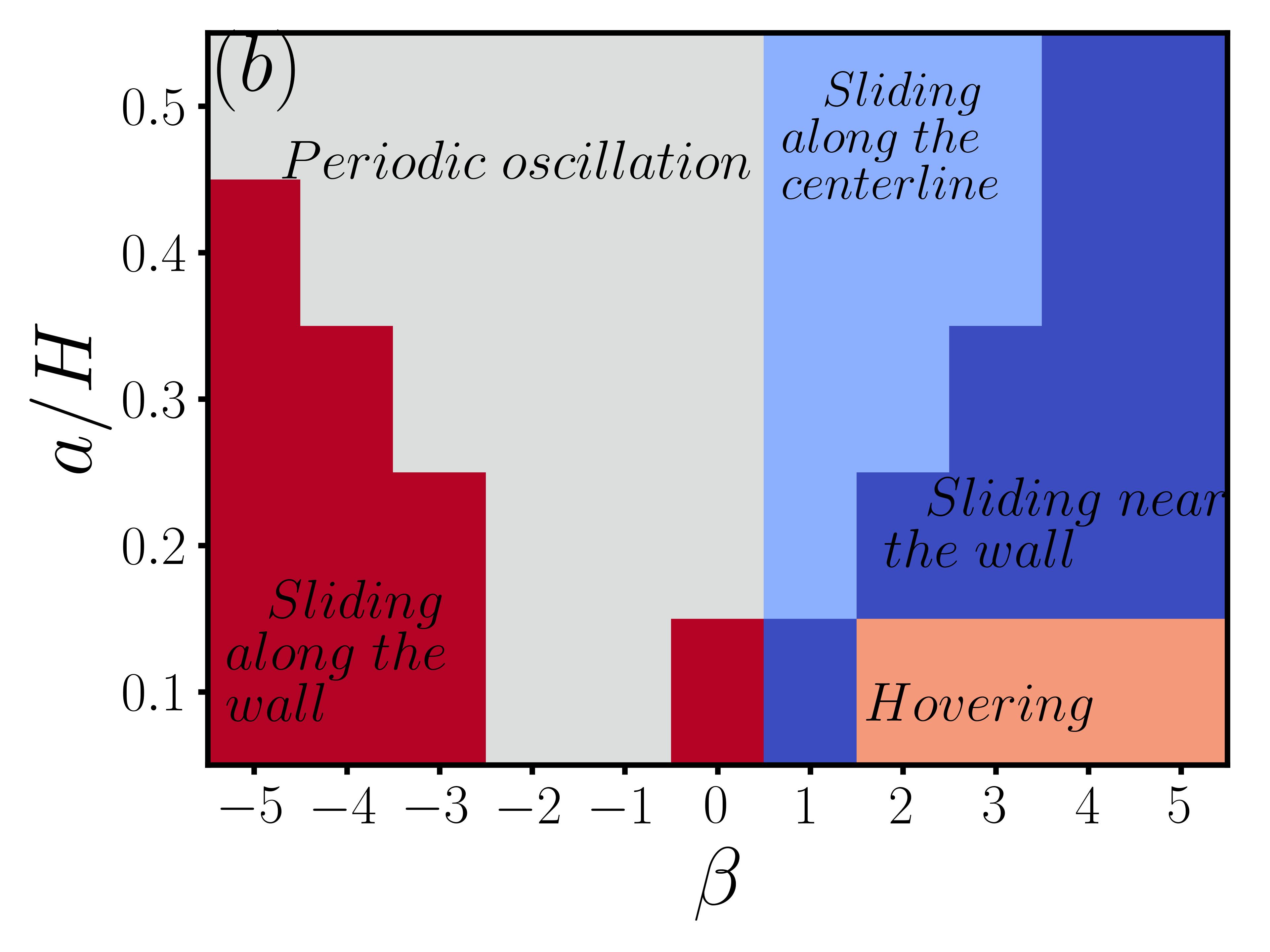}

\caption{Phase diagram in $\beta$ - $a/H$ plane showing different behaviour of squirmers in the mid plane of a square channel for two different initial conditions: (a) $y^{*}_{i}=0$,  $\theta_{i}=60^{0}$ and (b) $y^{*}_{i}=0$,  $\theta_{i}=30^{0}$.}
    \label{fig:phase}
\end{figure}

Fig.~\ref{fig:phase}(a) and (b) show that five kinds of swimming motions can be observed: sliding along the wall, near the wall, along the channel center-line, and then hovering and periodic oscillations.
Neutral swimmers mainly perform oscillatory motion except in very weak confinements ($a/H \lessapprox 0.1 $). Pushers perform both sliding along a wall and oscillatory motion depending on the values of $\beta$ and $a/H$. Stronger the $\beta$, larger the $a/H$ at which oscillatory behaviour emerges. 
At the same time, pullers exhibit hovering behavior, and sliding along the channel center-line or near a wall depending upon the confinement ratio $a/H$ and the initital condition.
Overall, on comparing the two phase diagrams in Fig.~\ref{fig:phase}, it is evident that initial conditions play an important role in deciding the trajectories of channel confined squirmers. 

In fact, it is clear from these phase diagrams that the sliding motion of pushers and hovering behavior of pullers occur for a larger range of confinement ratios as the strength of the squirmer ($\beta$) increases. This behavior is also consistent with the predictions of far-field approximations, as we can get the sliding motion of pushers and the hovering behavior of pullers as we increase the strength of force dipole in the far-field calculations.
%
%

%%%%%%%%%%%%%%%%%%%%%%%%%%%%%%%%%%%%%%%%%%%%%%%%%%%%%%%%%%%%%%%%%%%%%%%%%%%%%%%%%%%%%%%%%%%%%%%%%%%%%%%%%%%%%%%%%%%%%
\subsection{3D trajectories in square channels}
\label{subsec:3d_trajectory}
In this section, we will analyze the three-dimensional trajectories of squirmers in a square channel ($AR=0$). Similar to the findings in Fig.~\ref{fig:phase}, we see that the three dimensional trajectories are also dependent on the initial conditions. 
Therefore, in general, it is hard to categorize the trajectories, and we look for common and contrasting features of the trajectories of various swimmers as we discuss below. 
Such a discussion will give us interesting insights into the behaviour of channel confined microswimmers.
In the simulations reported here, the squirmers are initially placed in the $x-y$ plane, away from the channel center in the $z$-direction ($z^{*}_{i}=\pm 0.77$) with orientation $\theta_{i}$ ($-30^{0} \leq \theta_{i} \leq 30^{0}$), which gives the squirmer a three dimensional trajectory inside the channel.
%
%%%%%%%%%%%%%%%%%%%%%%%%%%%%%%%%%%%%%%%%%%%%%%%%%%%%%%%%%%%%%%%%%%%%%%%%%%%%%%%%%%%%%%%%%%%%%%%%%%%%%%%%%%%%%%%%%%%%%
\subsubsection{Helical trajectories of pushers}
\begin{figure*}
    \centering
\includegraphics[clip,trim=0cm 0cm 0cm 0cm,width=.328\textwidth]{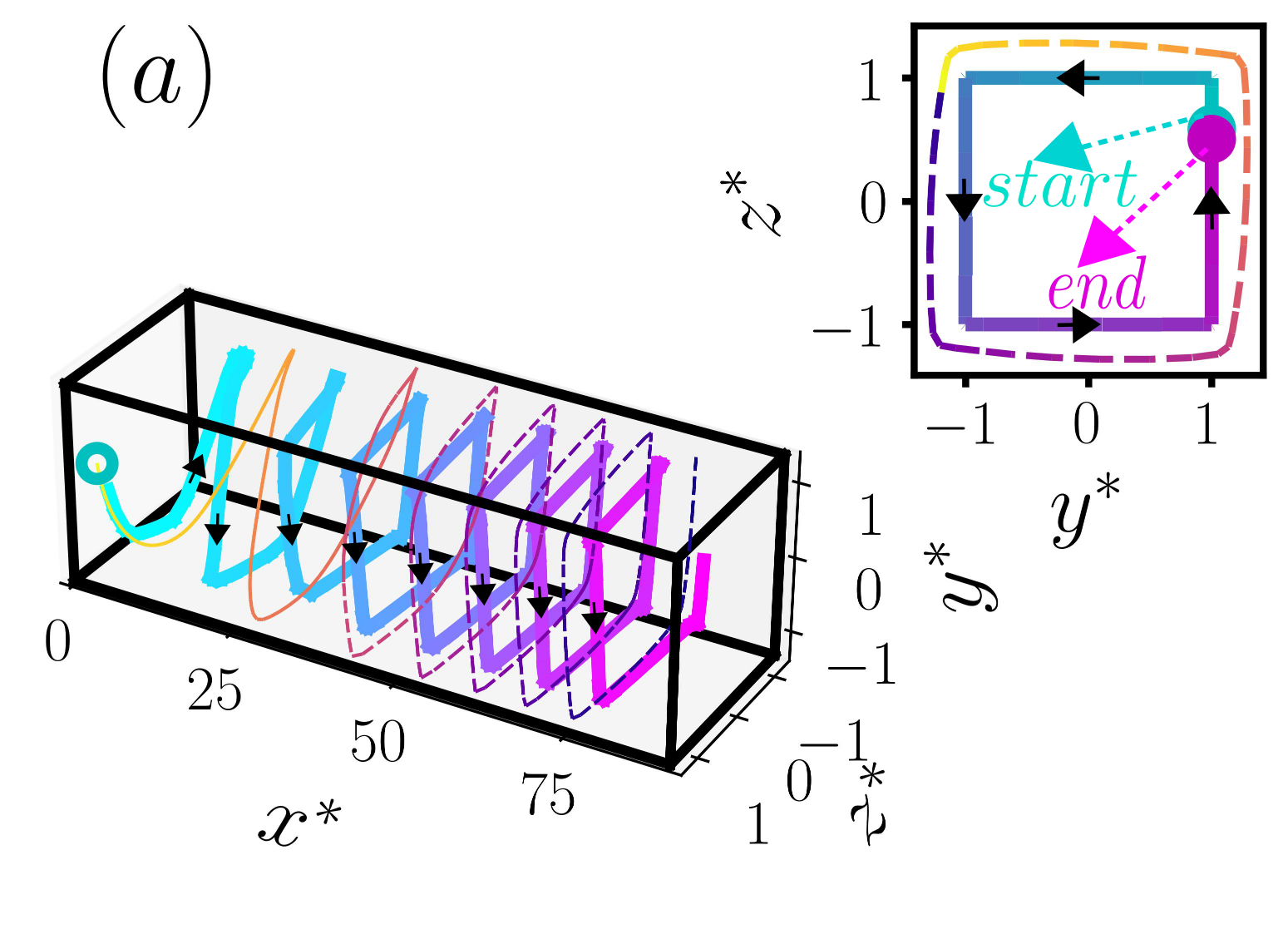}
  %\vspace{1cm}
  \includegraphics[clip,trim=0cm 0cm 0cm 0cm,width=.328\textwidth]{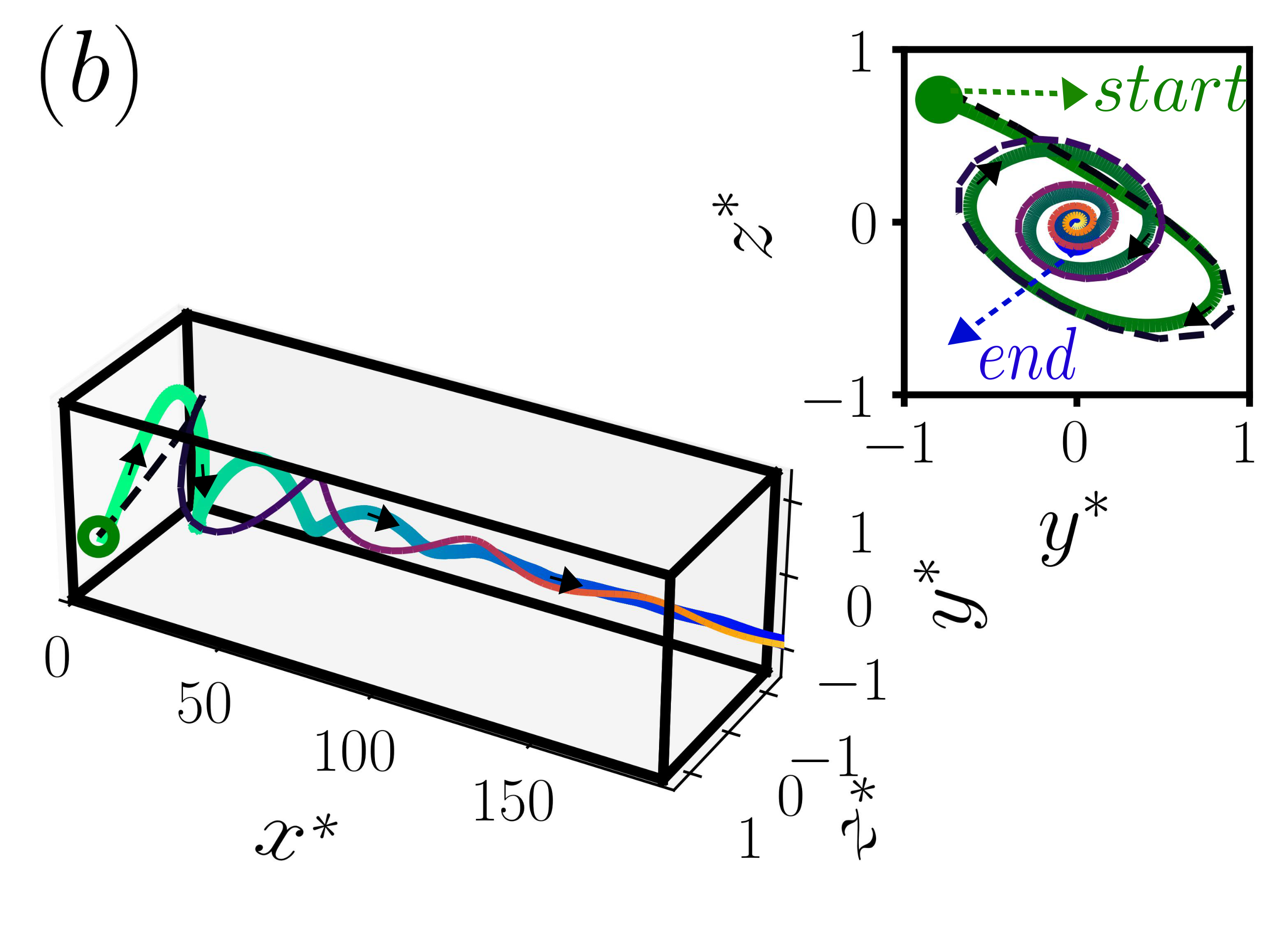}
            %\vspace{1cm}
          \includegraphics[clip,trim=0cm 0cm 0cm 0cm,width=.328\textwidth]{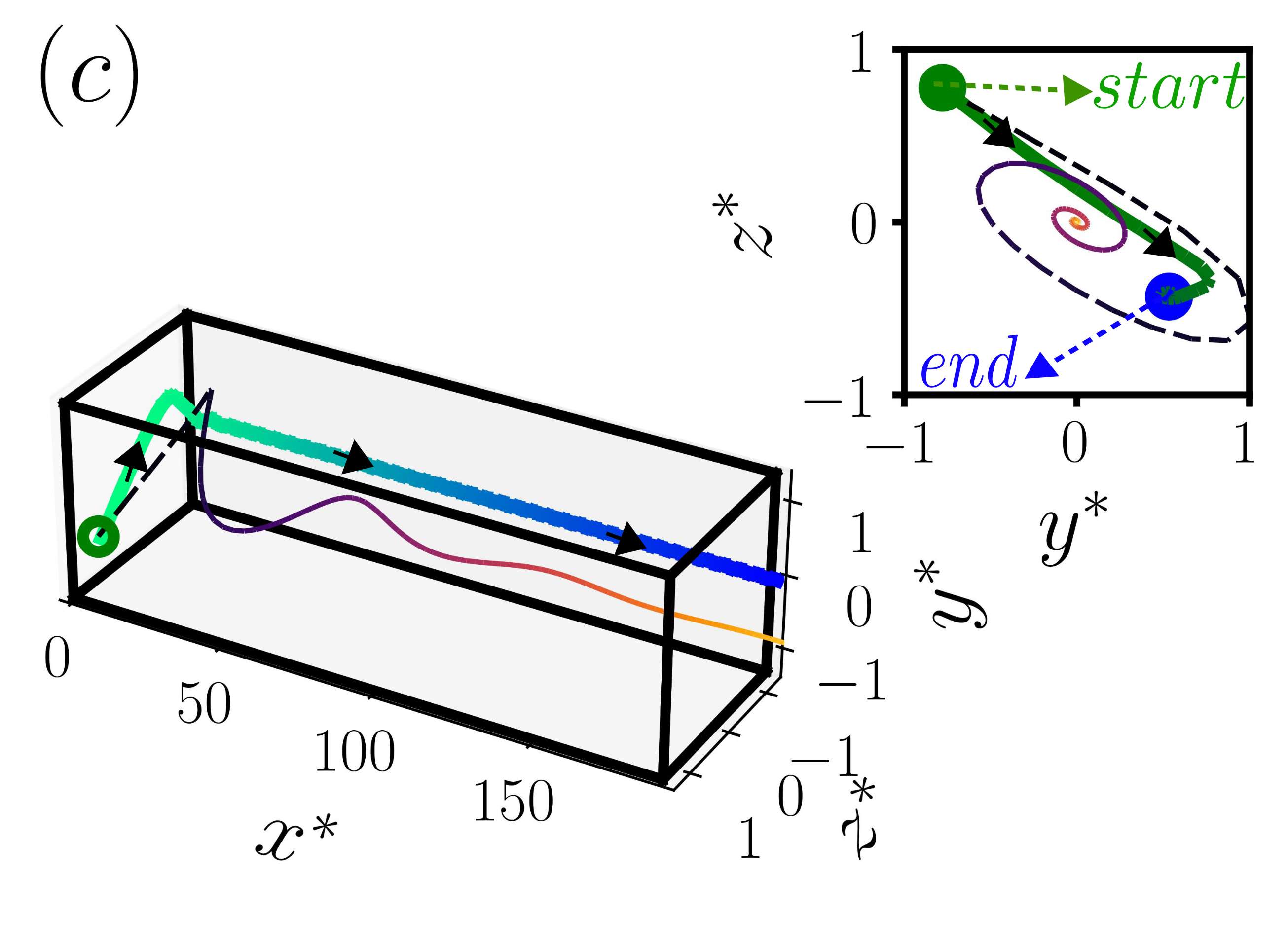}
            %\vspace{1cm}
            %\\
           \includegraphics[clip,trim=0cm 0cm 0cm 0cm,width=.338\textwidth]{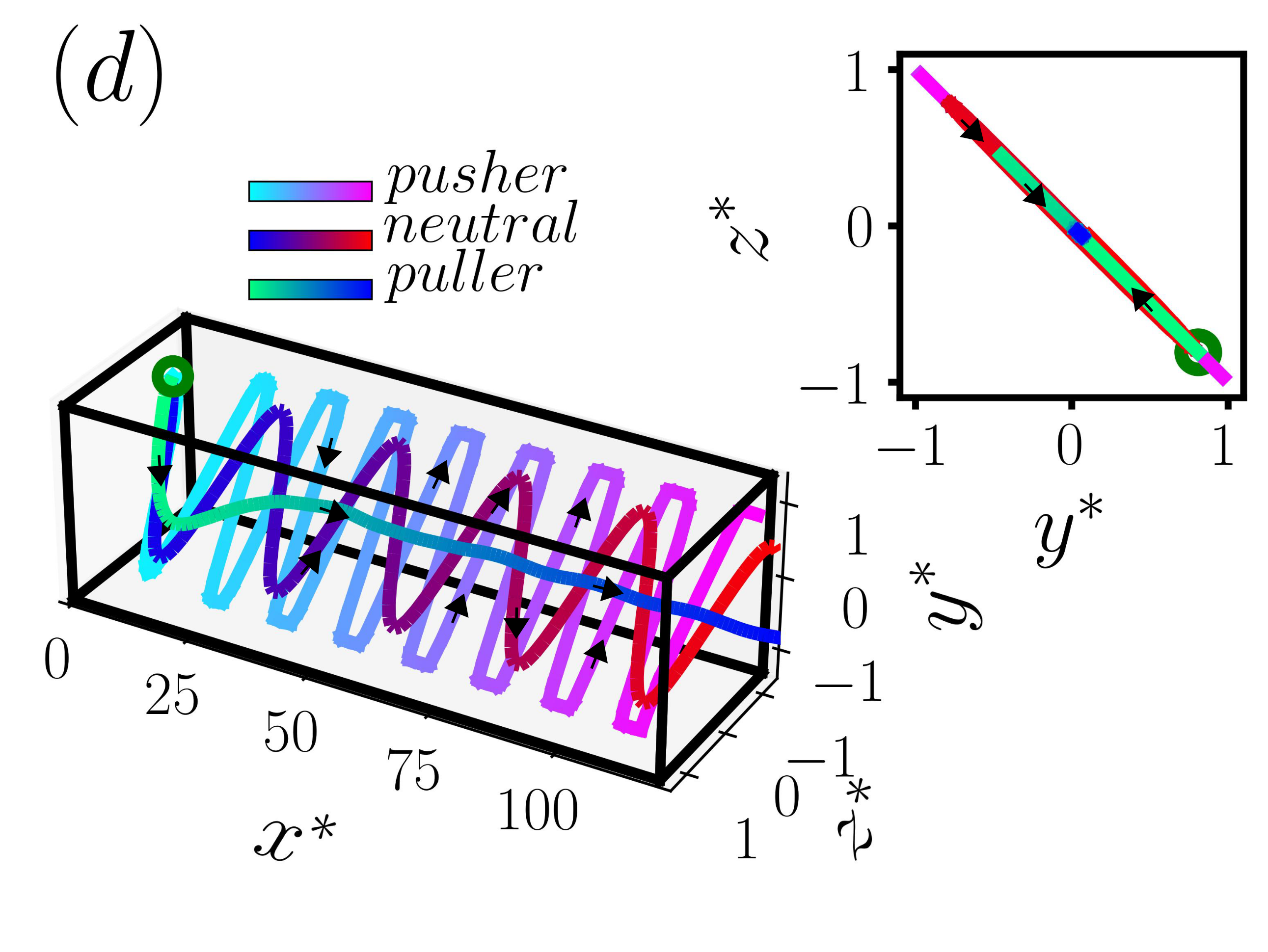}
           \includegraphics[clip,trim=0cm 0cm 0cm 0cm,width=.328\textwidth]{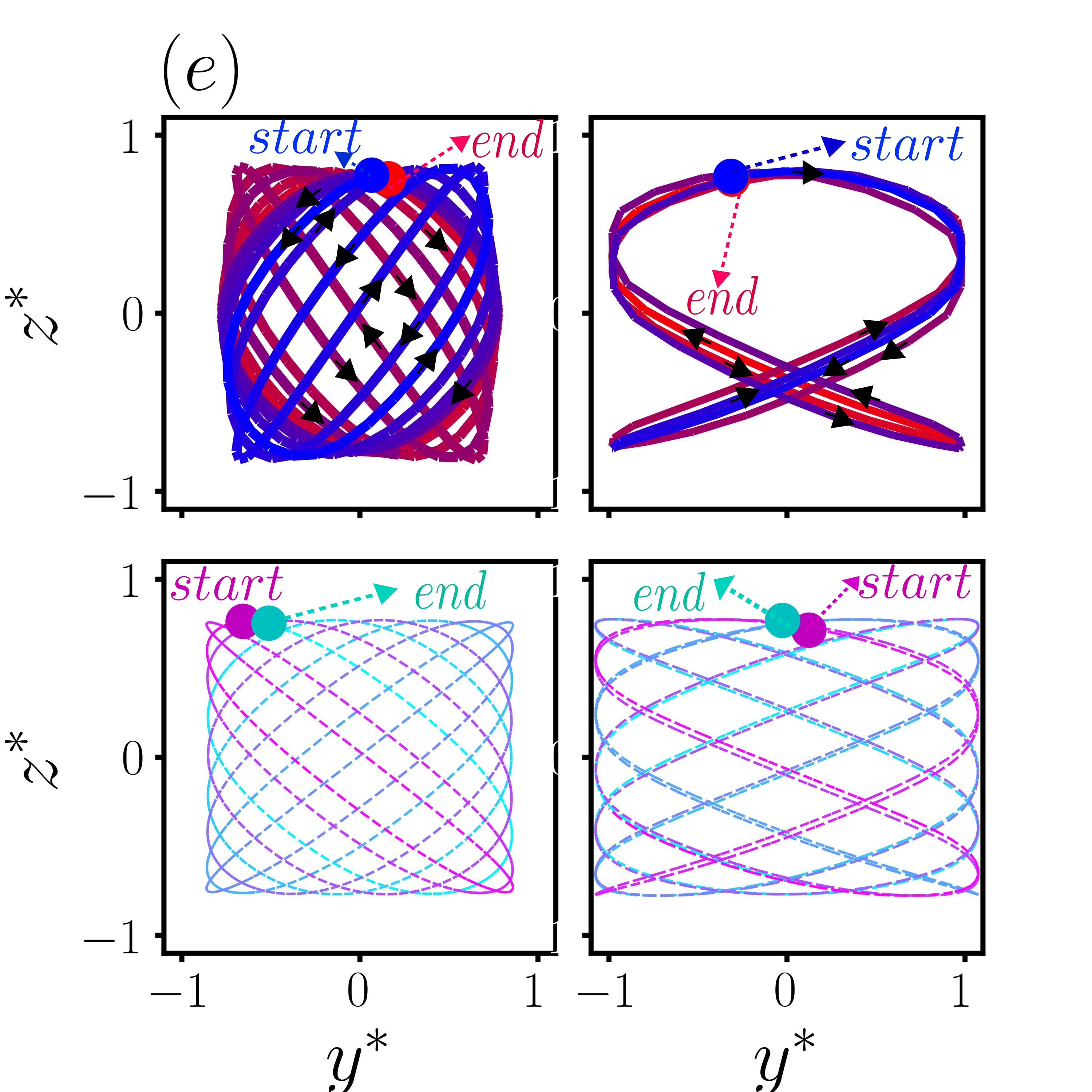}

\caption{The three-dimensional trajectory of micro-swimmers in a square channel (a) of the pusher with initial conditions $y^{*}_{i}=0.0$, $z^{*}_{i}=0.77$ and $\theta_{i}=15^{0}$, (b) of a weak puller ($\beta=1$) with initial conditions $y^{*}_{i}=-0.77$, $z^{*}_{i}=0.77$ and $\theta_{i}=10^{0}$,  (c) of a strong puller ($\beta=5$) with same initial conditions as in (b). The projected trajectory in the $y^{*}-z^{*}$ cross section is also shown (excluding the initial transients) on the right top of each figure. (d) The 3D trajectory of all three type of micro-swimmers with an initial condition $y^{*}_{i}=z^{*}_{i}=0.77$ and $\theta_{i}=0^{0}$, and they move in a plane that connect opposite edges of the channel. (e) The 3D trajectory of neutral swimmer projected to $y^{*}-z^{*}$ plane (excluding transients) for different initial conditions (top two figures) compared with corresponding far-field trajectories (bottom figure). The start and end points are also marked in the trajectory plots.}
    \label{fig:3d_pusher}
\end{figure*}

We will first study the case of pushers. We have already seen that pushers make oscillatory motion (Sec.~\ref{subsubsec:oscillation_pushers}) in the $x$-$y$ plane.
Similarly, in three dimensions, pushers exhibit a helical trajectory as illustrated in Fig.~\ref{fig:3d_pusher}(a). 
After the initial transience, pushers execute helical trajectories which span the entire channel.
As in two dimensions, here we find that the amplitude of these helical trajectories is independent of initial conditions. 
The two dimensional projections of these helical trajectories on $y^{*}-z^{*}$ plane is a closed curve as shown in the inset. 
The start and end points, and the direction of motion are also indicated in these figures. 
It can be seen that the helices conform to the inside surface of the channel, and hence the amplitude of the helix is determined by the channel size. 

An interesting variation of the aforementioned helical trajectories happens when the pusher is initially placed at a point that is equidistant in the $y$ and $z$-directions from the centerline ($y^*_{i} =z^*_{i}$), and with an orientation parallel to the $x$-axis ($\theta_{i} =0$).
In this position, the hydrodynamic interactions and the wall-induced velocities in the $y$ and $z$-direction will be equal in magnitude, and the pusher will exhibit an oscillation through a line connecting opposite edges, which we refer to as an edge-to-edge oscillation (Fig.~\ref{fig:3d_pusher}(d)).  

\begin{figure}
    \centering
    \includegraphics[clip,trim=0cm 0cm 0cm 0cm,width=.338\textwidth]{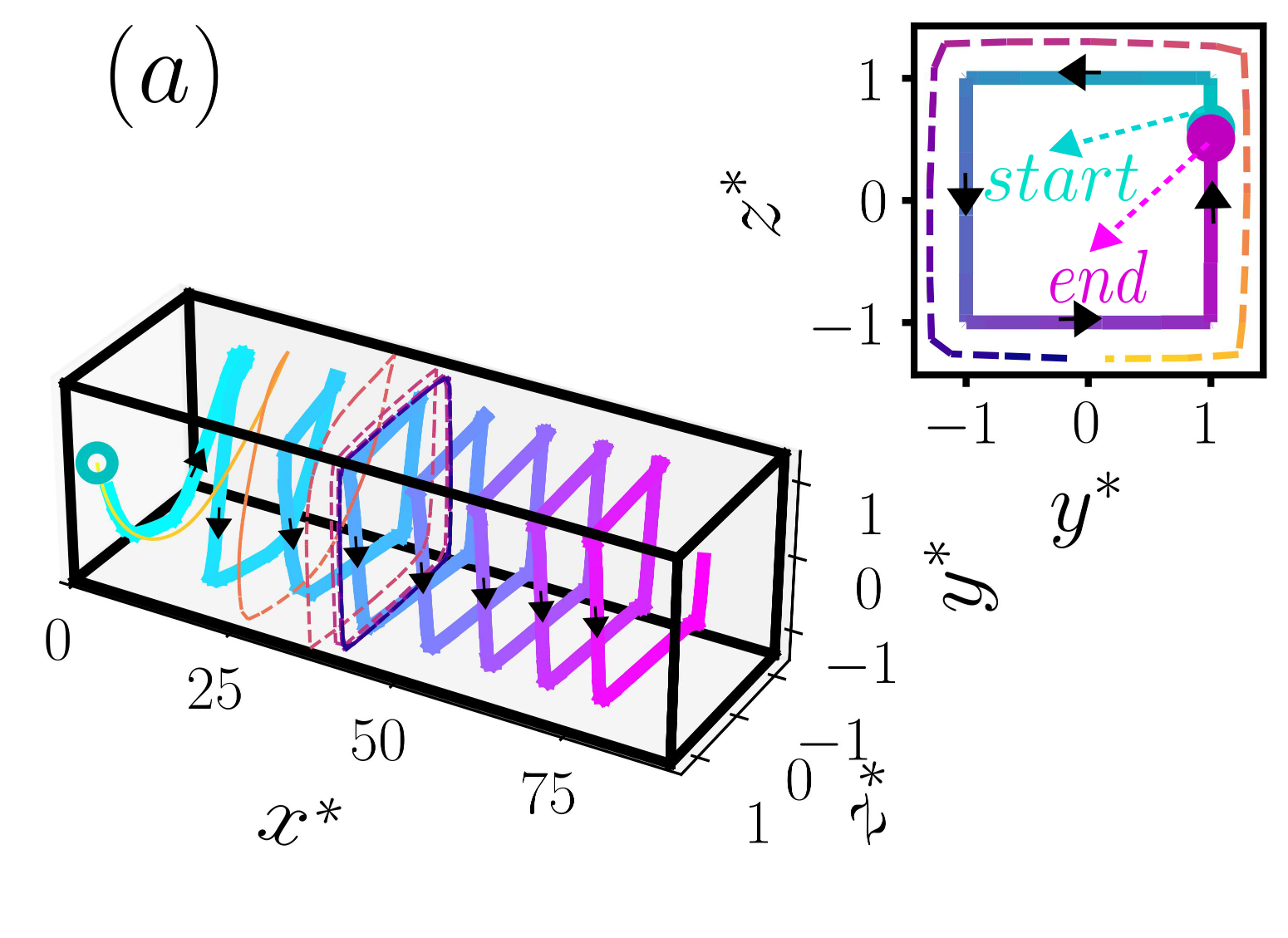}
           \includegraphics[clip,trim=0cm 0cm 0cm 0cm,width=.45\textwidth]{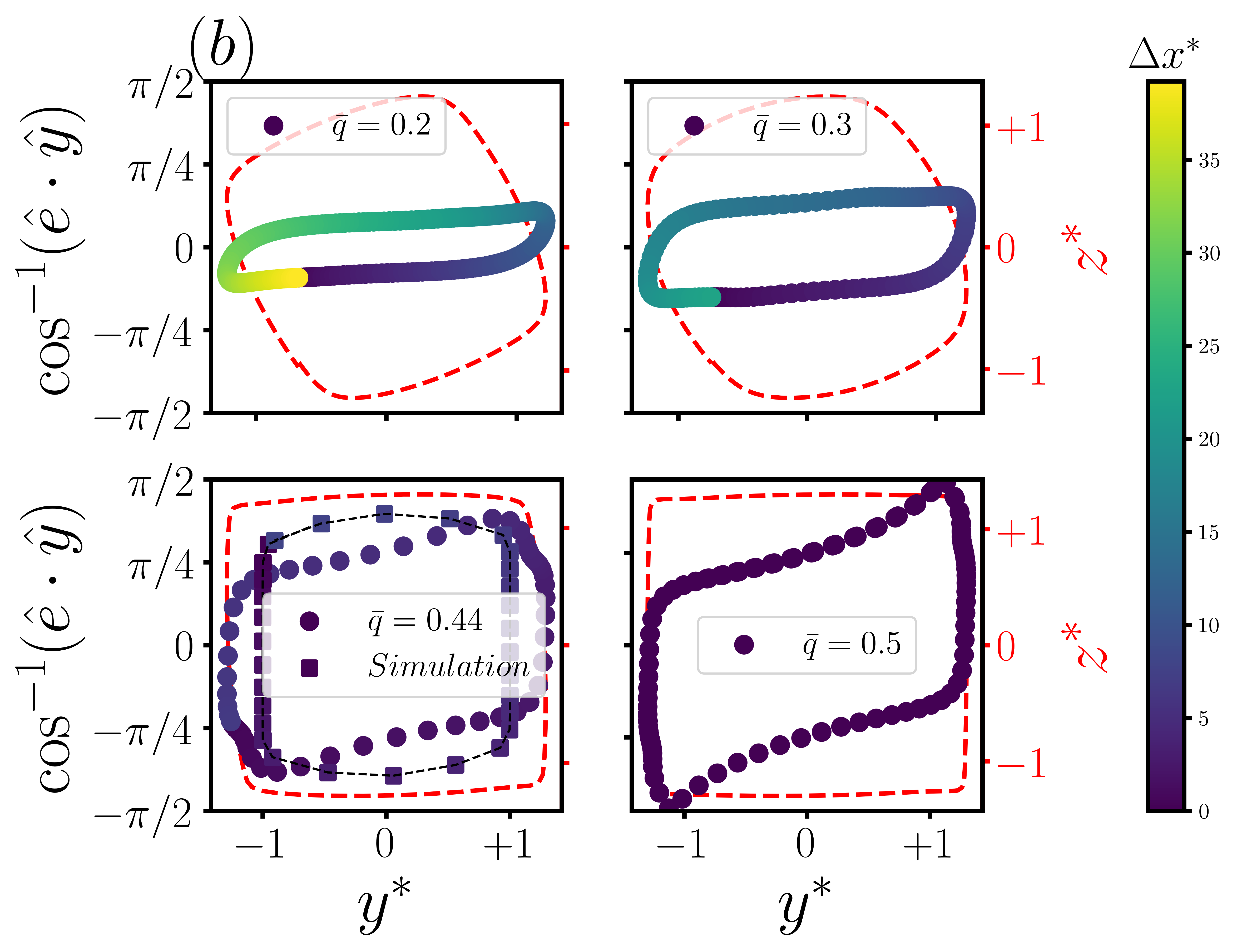}

\caption{(a) Comparison of the three-dimensional trajectory of pusher from simulation and far-field calculation (dotted line) for $\bar{q}=0.5$. The far-field calculations predict a steady state where pusher continuously orbits in the $y-z$ plane without moving along the $x$-axis. (b) The phase plots, from far-field calculations, showing the angle of orientation (angle between $\hat{\boldsymbol{e}}$ and the $y$-axis ) variation with  $y^{*}$ for various values of $\bar{q}$. The color bar represents the displacement in the $x$-direction while completing one cycle. The red dashed lines are the $y^{*}-z^{*}$ cross section.}
    \label{fig:3d_pusher_ff}
\end{figure}

The trajectories calculated using the far-field approximation are also shown (dashed lines) in Fig.~\ref{fig:3d_pusher}(a). 
These calculations are in good agreement with the simulation data. 
The $y^{*}-z^{*}$ plane of the trajectory is shown in the inset. 
Note that the far-field calculations show two different steady-state behavior for channel-confined pushers. 
The first one is the helical trajectory, as shown in Fig.~\ref{fig:3d_pusher}(a). 
The second one is the case wherein the pusher does not translate $x$-direction but it orbits the $y-z$ plane spanning the entire channel (dashed line in Fig.~\ref{fig:3d_pusher_ff}(a)). 
These steady state behaviours are dependent on the strength of the source dipole $\bar{q}$, for a given value of force dipole. 
The transition between these two steady state behaviours for increasing values of $\bar{q}$ are shown in the phase plots in Fig.~\ref{fig:3d_pusher_ff}(b).
The angle that the pusher makes with the $y$ axis is plotted against $y^{*}$, and the trajectories (stable limit cycles) are color coded with the displacement in the $x$-direction.
This displacement is zero for $\bar{q}=0.5$, and the phase plot corresponds to the steady state where the pusher continuously orbits in the $y-z$ plane without moving along the $x$-axis. 
When $\bar{q}$ is reduced to 0.44, the pusher exhibits finite displacement along the $x$-direction, while swimming in a helical trajectory.
Note that the results from the numerical simulation and the far-field calculations match quite well for  $\bar{q}=0.44$. Hence, we have considered $\bar{q}=0.44$ in our far-field calculations reported in Fig.~\ref{fig:3d_pusher}(a). 
With further reduction in $\bar{q}$, the pusher continues to execute the helical motion, but does not adhere to the contours of the channel cross-section (red dashed lines in Fig.~\ref{fig:3d_pusher_ff}(b)).
In summary, far-field calculations for pushers predict that for a particular force dipole strength, there is a critical source dipole strength above which the pusher does not move along the channel axis, and below which the pusher executes helical motion. However, it must be noted that numerical simulations did not show such a change in  behaviour for pushers.

%%%%%%%%%%%%%%%%%%%%%%%%%%%%%%%%%%%%%%%%%%%%%%%%%%%%%%%%%%%%%%%%%%%%%%%%%%%%%%%%%%%%%%%%%%%%%%%%%%%%%%%%%%%%%%%%%%%%%
\subsubsection{3D motion of neutral swimmers}
\label{sec:3d motion of neutral swimmers}
We have already seen that neutral swimmers execute oscillatory trajectories in two dimensions, and the wavelength and amplitude of these trajectories are dependent on initial conditions. 
Similarly, in three dimensions, neutral swimmers exhibit initial condition dependent, complex trajectories. Examples of such trajectories, projected onto the $y^{*}-z^{*}$ plane, are shown in Fig.~\ref{fig:3d_pusher}(e). 
In the top-left figure shown in Fig.~\ref{fig:3d_pusher}(e), the initial conditions are $y^{*}_{i}=0.11$, $z^{*}_{i}=-0.77$ and $\theta_{i}=15^{0}$.
In this case, the  neutral squirmer follows a clockwise trajectory followed by an anticlockwise trajectory while translating along the channel length thus making an aperiodic helical trajectory. 
In the top-right example of Fig.~\ref{fig:3d_pusher}(e), the initial conditions are ($y^{*}_{i}=0$, $z^{*}_{i}=0.77$ and $\theta_{i}=-30^{0}$). 
In this case, the neutral swimmer exhibits a ribbon like pattern in the $y^{*}- z^{*}$ plane. 
Such initial condition dependency, and the variety of the trajectories, make it hard to characterise the trajectories of channel confined neutral swimmers.
Nevertheless, the origin of this behaviour will be discussed later in Sec.~\ref{subsubsec:3d_as_combination}. 
Note, the initial conditions $y^{*}_{i}=z^{*}_{i}=0.77$ and $\theta_{i}=0$ would give rise to a edge-to-edge oscillating trajectory (Fig.~\ref{fig:3d_pusher}(d)) as we saw in the case of pushers.

The bottom panel of Fig.~\ref{fig:3d_pusher}(e) shows the far-field trajectory of the corresponding simulation in the top panel. 
In this case, the far-field trajectories (dashed lines) may look different from the results from the numerical simulations. As we will see, the superposition principle described in Sec.~\ref{subsubsec:3d_as_combination} will explain this difference observed from the results of two methods. 

%%%%%%%%%%%%%%%%%%%%%%%%%%%%%%%%%%%%%%%%%%%%%%%%%%%%%%%%%%%%%%%%%%%%%%%%%%%%%%%%%%%%%%%%%%%%%%%%%%%%%%%%%%%%%%%%%%%%%
\subsubsection{Sliding motion of pullers along or parallel to the channel centerline}
Fig.~\ref{fig:3d_pusher}(b) and (c) show the three-dimensional trajectories of pullers and their projections onto the $y^{*}-z^{*}$ plane. 
Trajectories in both figures correspond to same initial conditions $y^{*}_{i}=-0.77$, $z^{*}_{i}=0.77$ and $\theta_{i}=10$, but with increasing strength of the puller. 
For a weak puller (Fig.~\ref{fig:3d_pusher}(b)), the trajectory exhibits a spiralling motion initially but  the amplitude of the spiral decreases eventually and the puller slides along the channel centerline.
However, for a stronger puller (Fig.~\ref{fig:3d_pusher}(c)), the sliding motion is closer and parallel to one of the walls.
These results are also consistent with the sliding motion of pullers observed in two-dimensions (discussed in Sec.~\ref{subsubsec:sliding_pullers}). 

The far-field calculations (dashed lines) predict a comparable trajectory for a weak puller, $\beta=1$ (Fig.~\ref{fig:3d_pusher}(b)).
But, as in the 2D case, the far-field calculations fail to predict the sliding motion near a wall for strong pullers in 3D.

%%%%%%%%%%%%%%%%%%%%%%%%%%%%%%%%%%%%%%%%%%%%%%%%%%%%%%%%%%%%%%%%%%%%%%%%%%%%%%%%%%%%%%%%%%%%%%%%%%%%%%%%%%%%%%%%%%%%%
\subsubsection{Understanding 3D trajectories as a superposition of 2D trajectories}
\label{subsubsec:3d_as_combination}
In this section, we introduce a simple technique to understand and predict the 3D trajectories of a channel confined squirmer. 
The idea is to utilize the results obtained in the previous sections, specifically the trajectories in two-dimensions, and to use the principle of superposition to construct the behaviour of channel confined swimmers in three-dimensions. 

Let us first consider the two-dimensional trajectory of a squirmer confined to $x-y$ plane of the channel. 
In general, the trajectory can be mathematically expressed as
\begin{align}
    y(x) = A \sin(\frac{2\pi}{ \lambda}x+\Delta)\exp(-m x)
    \label{eqn:supxy}
\end{align}
where $A$ is the amplitude, $\lambda$ is the wavelength, $\Delta$ is the phase of the oscillations, and $m$ is a decay constant. 
The steady state trajectories of pushers and neutral swimmers are oscillatory and periodic, which can be represented by this general expression with $m = 0$. 
Transient path of pullers which follow an exponential decay, but settle to a sliding trajectory can be represented with $ m \ne 0$. 
For pullers, the initial conditions determine the exact value of $m$. 
Similarly, for neutral swimmers $A$ and $\lambda$ are also determined by initial conditions.

An identical expression for the 2D trajectory of a squirmer in the $x-z$ plane can be written as
\begin{align}
z(x) = \tilde{A} \sin(\frac{2\pi}{ \tilde{\lambda}}x)\exp(-\tilde{m}x).
\label{eqn:supxz}
\end{align}
where $\tilde{A}$, $\tilde{\lambda}$ and $\tilde{m}$ have the same meaning as given above, and their values are determined by the squirmer parameters, the channel dimensions, and the initial conditions.

We propose that the three-dimensional trajectory of a squirmer confined in a channel can be constructed by assuming a superposition of the two-dimensional motions in the two orthogonal planes, namely $x-y$ and $x-z$ planes. 
Therefore, the two dimensional position vector in the $y-z$ plane, $\mathbf{T}_{s}(x) = y(x) \hat{y} + z(x) \hat{z}$, describing the trajectory of the squirmer as a function of $x$ can be expressed as,
\begin{equation}
    \begin{split}
\mathbf{T}_{s}(x) =& A \sin(\frac{2\pi}{ \lambda}x+\Delta) \exp(-mx) \hat{y} \\
&+ \tilde{A} \sin(\frac{2\pi}{ \tilde{\lambda}}x)\exp(-\tilde{m}x)  \hat{z} .
\label{eqn:superposition}
    \end{split}
\end{equation}
Thus, the assumption of superposition yields the above expression to determine the trajectory of the squirmer in three-dimensions from its two-dimensional counter parts. 
We now analyze the trajectory obtained from Eq.~\ref{eqn:superposition} with the actual trajectories.

\noindent\textit{Pushers:} The amplitude and wavelength of two-dimensional trajectories of pushers are independent of initial conditions. 
Therefore, in channels of square cross section $\tilde{A}=A$ and  $\tilde{\lambda}=\lambda$.
Eliminating $x$ from Eq.~\ref{eqn:superposition}, we can derive an expression for the motion of the squirmer in $y-z$ plane as,
\begin{align}
\frac{z^{2}}{A^{2}}-\frac{2zy}{A^{2}}\cos\Delta+\frac{y^{2}}{A^{2}}=\sin^{2}\Delta
\label{eqn:ellipse}
\end{align}
which is the general equation for an ellipse.
Eq.~\ref{eqn:ellipse} implies that the 3D trajectory of a channel confined pusher will be a helix.
The shape of the helix will be determined by the parameter $\Delta$, the phase difference between the waves in the $x-y$ and the $x-z$ planes. 
For $\Delta=0$, when there is no phase difference between the two-dimensional trajectories considered in the two orthogonal planes, Eq.~\ref{eqn:ellipse} reduces to the equation of a straight line $y = z$.
This case corresponds to the edge to edge oscillations seen for a pusher. 
For $\Delta \ne 0$, the phase difference between the waves results in the three-dimensional helical trajectory of the pushers (Fig.~\ref{fig:3d_pusher}(a)).

\begin{figure}
    \centering
 \includegraphics[clip,trim=0cm 0cm 0cm 0cm,width=.238\textwidth]{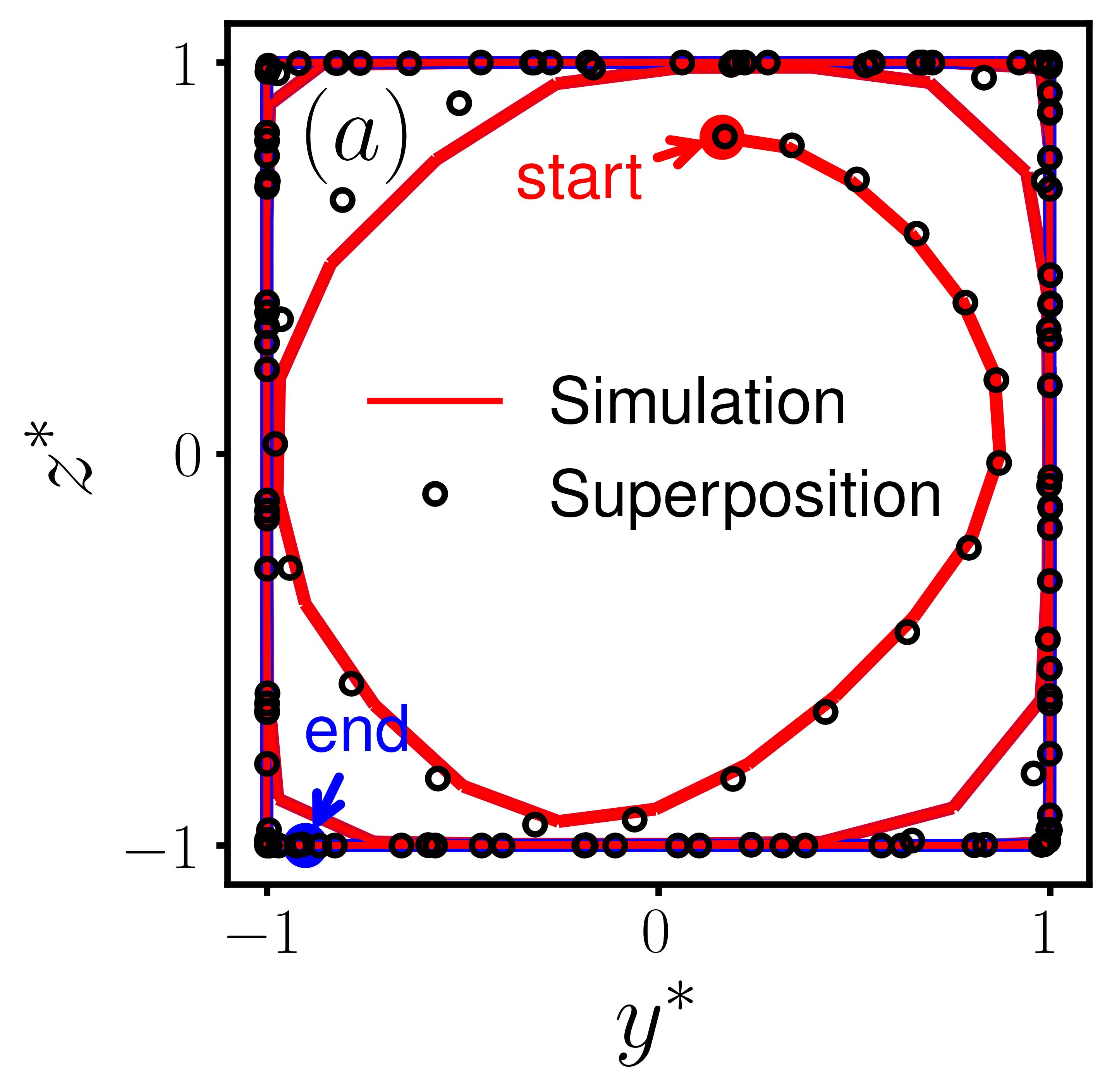}
  %\vspace{1cm}
          \includegraphics[clip,trim=0cm 0cm 0cm 0cm,width=.238\textwidth]{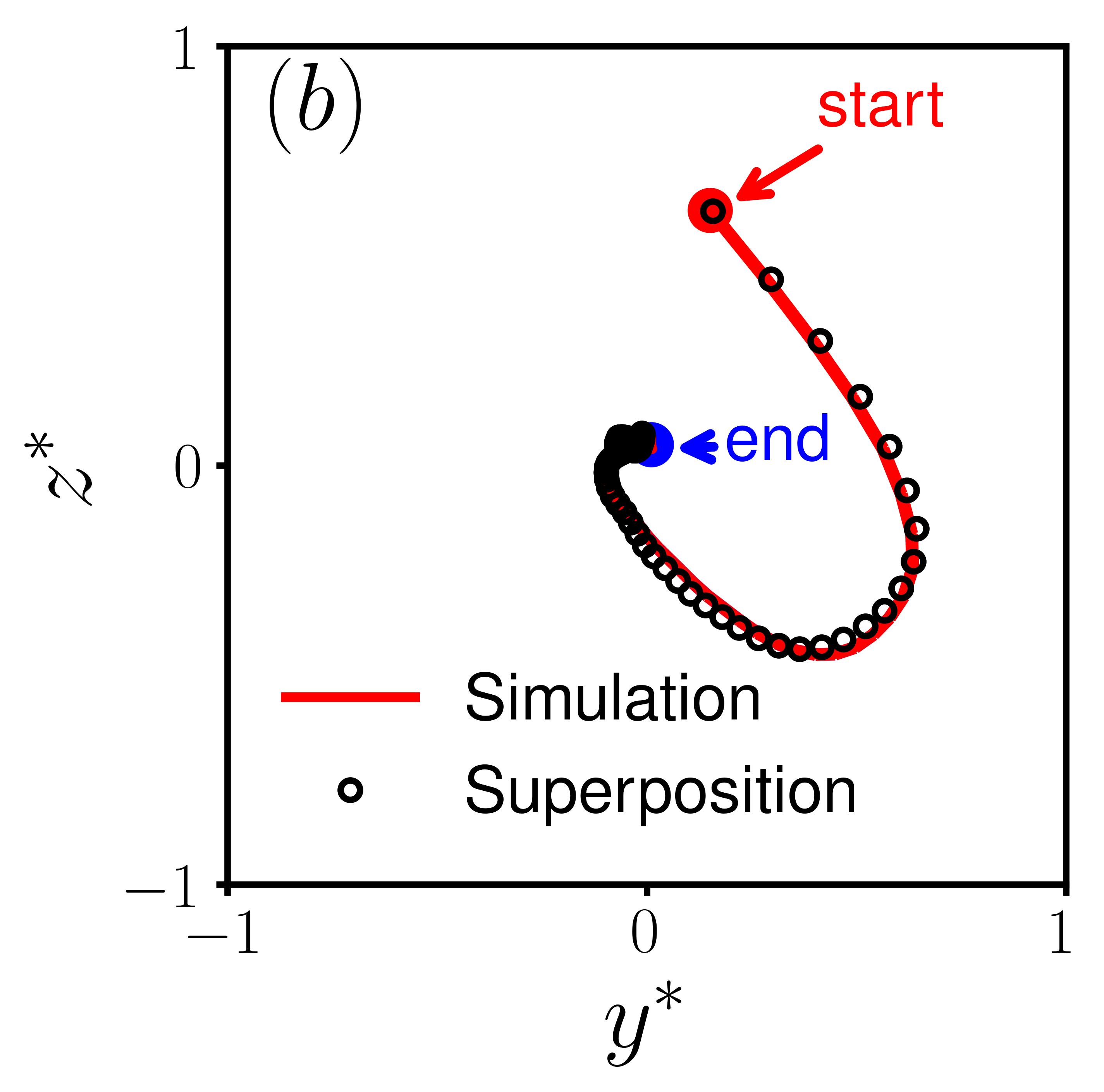}
            %\vspace{1cm}

\caption{Comparison of the trajectories (projected on to $y^* - z^*$ plane) resulting from full-scale 3D simulation and from the superposition of 2D trajectories in the $x-y$ and $x-z$ planes (a) for pushers and (b) for pullers.  In both (a) and (b) the initial conditions are $y^{*}_{i}=0.0$, $z^{*}_{i}=0.77$, and $\theta_{i}=15^{0}$.}
    \label{fig:comparison}
\end{figure}

In the above discussion, we approximated the-two dimensional, oscillatory motion of pushers as a sinusoidal function to qualitatively explain the behaviour in three-dimensions.
Instead of these expressions (Eq.~\ref{eqn:supxy}-\ref{eqn:supxz}), we can superpose the actual two dimensional trajectories obtained from the simulations to perform the superposition. 
Succinctly put, the trajectories of the squirmer when restricted (by symmetry) to move in the $x-y$ and $x-z$ planes in a square channel obtained from simulations can be superposed to construct the corresponding three-dimensional trajectory, without actually solving for the 3D geometry.  
An example of such a result is plotted in Fig.~\ref{fig:comparison}(a). 
The trajectory from the full-scale 3D simulation (red line) is also shown in this figure. 
The agreement between the two approaches is quite good, indicating that the superposition principle works in generating three dimensional trajectories of pushers. 
More importantly, the principle of superposition explains the origin of the nature of the three-dimensional trajectory. 

\noindent\textit{Pullers:} A similar comparison between the 3D trajectories from the full scale numerical simulation and the superposition method for pullers also illustrates the efficacy of the latter.
For example, consider the comparison  made for pullers in a 3D channel with initial condition, $y_i^{*}=0.0$, $z_i^{*}=0.77$, and $\theta_i=15^{0}$. 
We can split this problem into the superposition of two 2D cases- first one in $x-y$ plane with  $y_i^{*}=0.0$, $z_i^{*}=0.0$, $\theta_i=15^{0}$ which isolates the effects of top and bottom wall, and the second one in $x-z$  plane with initial conditions $y_i^{*}=0.0$, $z_i^{*}=0.77$, and $\theta_i=0^{0}$ which isolates the effects of the side walls. 
Using the results of these two simulations, a three dimensional trajectory is constructed in Fig.~\ref{fig:comparison}(b).
This figure highlights the fact that the transient motion of pullers can be also understood as the superposition of two mutually perpendicular damping oscillations. 

\noindent\textit{Neutral swimmers:} We have already seen that the initial conditions affect 
the wavelength and amplitude of the oscillatory trajectory of neutral swimmers even in two dimensions.
Therefore, the trajectories for a neutral swimmer obtained in the $x-y$ and $x-z$ plane can have different wavelengths and amplitude. More over they differ in the phase too. Thus, the trajectory of a neutral swimmer can be understood as the superposition of two mutually perpendicular waves with different wavelengths and amplitudes. Naturally, the resultant trajectory will be complex  as illustrated in Fig.~\ref{fig:3d_pusher}(e) (Sec.~\ref{sec:3d motion of neutral swimmers}). However, interestingly, there can be an exceptional initial condition for which the amplitude, wavelength, and phase of oscillatory motion in $x-y$ and $x-z$ planes exactly match, \textit{i.e.,} $A=\tilde{A}$, $\lambda=\tilde{\lambda}$, $\Delta=0$, and hence, we get a edge-to-edge oscillation (Fig.~\ref{fig:3d_pusher}(d)) as we reported before.

Further, this analysis also allows us to comprehend the mismatch in the predictions of numerical simulations and far field calculations reported in Fig.~\ref{fig:3d_pusher}(e) (Sec.~\ref{sec:3d motion of neutral swimmers}). As mentioned above, in 2D, a neutral swimmer performs oscillations whose amplitude and wavelength depend on initial conditions. 
The far-field calculations predict 2D oscillations with slightly different amplitudes and wavelengths. 
Such differences in amplitude and wavelength may create a noticeable change while constructing the 3D trajectory as can be seen in Fig.~\ref{fig:3d_pusher}(e).

In summary, the superposition principle constructs the three-dimensional trajectories of channel confined pushers, pullers, and neutral swimmers. 
The match obtained between the results of the numerical simulation and that from the superposition provides an insight into the dominant effects that drive the dynamics of a squirmer in a three-dimensional channel.
If the motion of the squirmer is analyzed from the perspective of method of reflections, the net effect due to the four walls of the channel can be considered as the sum of (i) hydrodynamic interaction with the channel walls parallel to $x-y$ plane, (ii) hydrodynamic interaction with the channel walls parallel to the $x-z$ plane and (iii) the consequent series of reflections from the two pairs of walls in order to simultaneously satisfy the boundary conditions at all the walls.
Clearly, the above results illustrate that the effect of the last contribution, i.e. the consequent reflections, is much weaker and the principle of superposition calculates the approximate trajectory of the squirmer in three-dimensions. 
Although this method is not exact, it is sufficient to explain the origin of sliding motion of pullers and the helical motion of pushers in a square channel. 
This method also helps to reason the complex trajectory of neutral swimmers.

%%%%%%%%%%%%%%%%%%%%%%%%%%%%%%%%%%%%%%%%%%%%%%%%%%%%%%%%%%%%%%%%%%%%%%%%%%%%%%%%%%%%%%%%%%%%%%%%%%%%%%%%%%%%%%%%%%%%%
\subsection{3D motion of squirmers in rectangular channels}
\label{subsec:3d_rectangle}
\begin{figure*}
    \centering
\includegraphics[clip,trim=0cm 0cm 0cm 0cm,width=.252\textwidth]{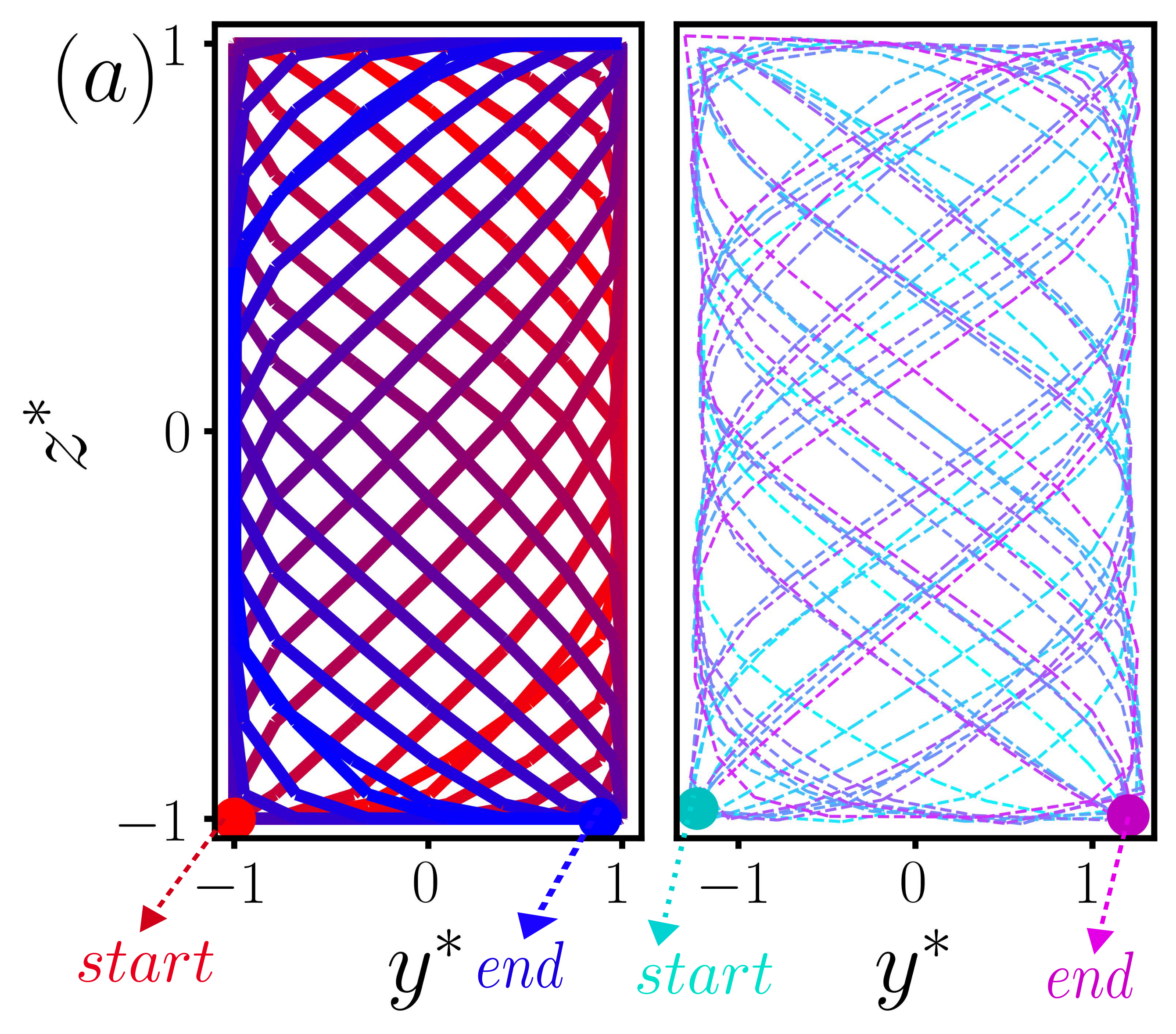}
  %\vspace{1cm}
          \includegraphics[clip,trim=0cm 0cm 0cm 0cm,width=.232\textwidth]{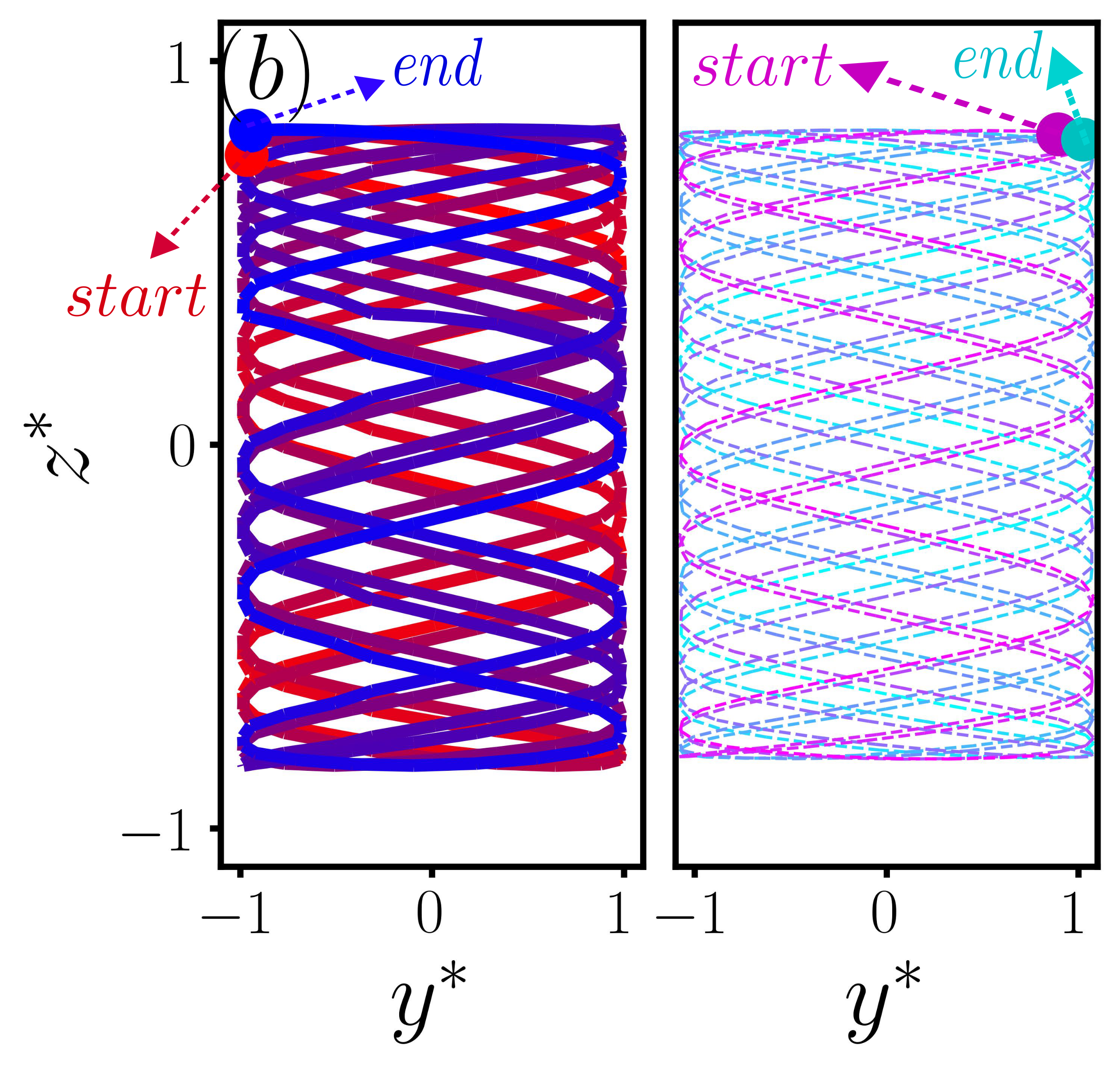}
          \includegraphics[clip,trim=0cm 0cm 0cm 0cm,width=.15\textwidth]{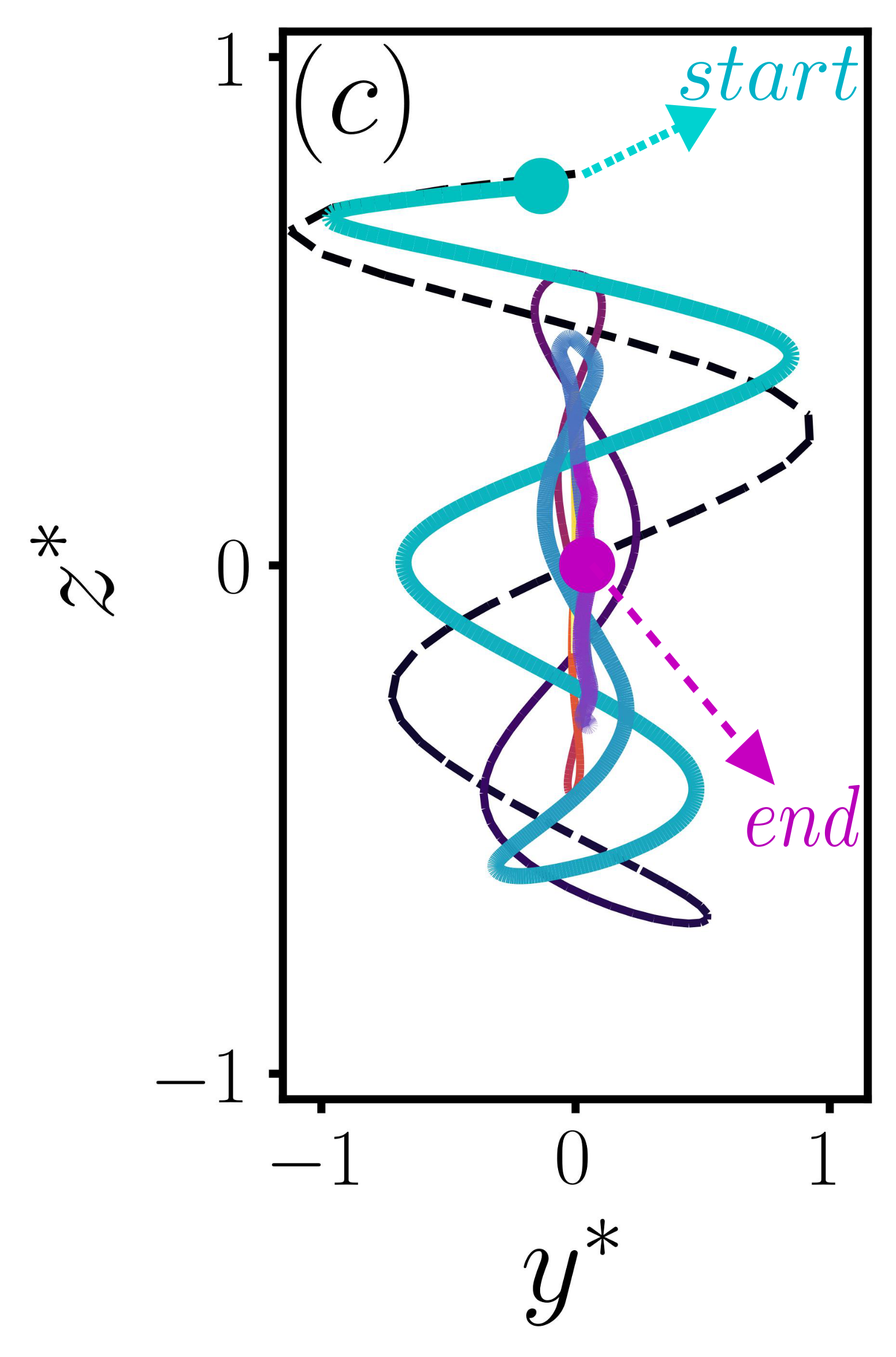}
          \includegraphics[clip,trim=0cm 0cm 0cm 0cm,width=.35\textwidth]{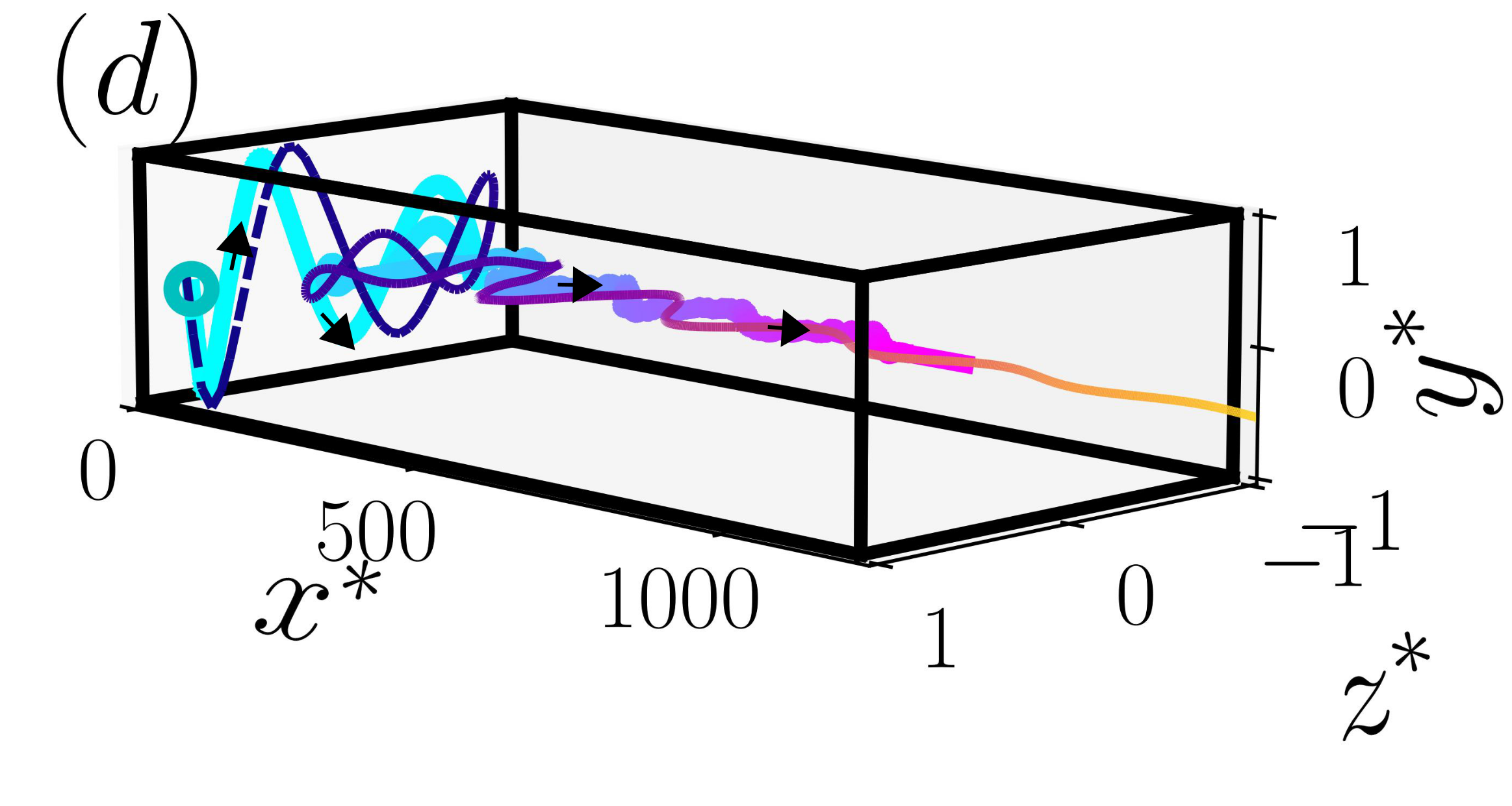}
            %\vspace{1cm}

\caption{Trajectories of a microswimmer in a 3D rectangular channel of $AR=0.5$. Projection of the trajectory onto the $y^{*}-z^{*}$ plane for (a) a  pusher, (b) a neutral swimmer and (c) a puller. Panel (d) shows the 3D trajectory of the puller reported in (c). The initial conditions for the swimmers are $z^{*}_{i}=0.77$ and $y^{*}_{i}=0.0$, $\theta_{i}=-30^{0}$. In (b) $z^{*}_{i}=0.66$.  The dashed lines in each figure are trajectories from the far-field calculations.}
    \label{fig:ar.5}
\end{figure*}
We conclude the study by extending our trajectory analysis to the case of squirmers in rectangular channels ($AR>0$). 
We carried out simulations by keeping the height of the channel fixed, while increasing the width of the channel systematically. 
The trajectories of the squirmers appeared complex as we increased the aspect ratio $AR$. 
Note that $AR = 0$ for a channel with square cross section (Eq.~\ref{eq:ar}).  
In this section, we report the  trajectories of different types of squirmers in a rectangular channel of $AR=0.5$, and explain the origin of the apparent complexity of the trajectory using the superposition principle.

We start by analysing the case of pushers.
The trajectories of pushers in this case are not easy to generalize since they do not follow the simple helical trajectories as observed in a square channel. 
Instead the trajectory appears complex, and these 3D trajectories strongly depend on the confinement ratio and aspect ratio of the channel. 
In Fig.~\ref{fig:ar.5}(a), the trajectory of a pusher, initially located at $y^{*}_{i}=0$, $z^{*}_{i}=0.66$ and oriented at $\theta_{i}=-30^{0}$, is shown in the $y^{*}-z^{*}$ plane. 
The left plot shows the simulation result, and the right plot with dashed lines is the far-field calculation. The trajectory is plotted with a color gradient for easy visualisation of the trajectory.
This seemingly complex trajectory can, again, be explained from the superposition method as follows. 
We have seen that the wavelength of the trajectory of a pusher increases with decreasing the confinement ratio, $a/H$ (Fig.~\ref{fig:Pushers_oscillation}(c)). 
In a rectangular channel, the width of the channel is more than the height. 
Therefore both the amplitudes and the wavelengths of the trajectories produced in the planar trajectories belonging to $x-y$ and $x-z$ planes will be different, \textit{i.e.,} $A \neq \tilde{A}$ and $\lambda \neq \tilde{\lambda}$.
The superposition of these two different waves gives rise to the appearance of the complex three-dimensional trajectory of the pusher shown in Fig.~\ref{fig:ar.5}(a). 

Next, we discuss the trajectories of pullers. 
They exhibit sliding motion in a rectangular channel as well.
As discussed earlier, the sliding can be either along the channel centerline or close to one of the walls. Moreover, the initial location and orientation dictate the  path to reach the final trajectory.  
Consider the case of a puller initially located at $y^{*}_{i}=0$, $z^{*}_{i}=0.66$ and oriented at $\theta_{i}=-30^{0}$ in a rectangular channel. 
The $y^{*}-z^{*}$ plane view, and the three dimensional trajectory followed by the puller is plotted in  Fig.~\ref{fig:ar.5}(c) and Fig.~\ref{fig:ar.5}(d) respectively. 
This swimmer eventually slides along the channel centerline for which $y^{*}=0$ and $z^{*}=0$. 
The transient path that leads to this final state can be analyzed as follows. 
During the motion, the puller first reaches the $y^{*}=0$ plane, namely the center of the $x-y$ plane. 
Then it moves only along the $z$-direction to the point $z^{*}=0$, thus reaching the channel centerline. 
Hence, the entire trajectory of the puller can be understood as a combination of two individual trajectories, consistent with the principle of superposition. 

In rectangular channels, as we increase $AR$, neutral swimmers also generate quite complicated trajectories.
An example of the trajectory of a neutral swimmer in $y^{*}-z^{*}$ plane is shown in Fig.~\ref{fig:ar.5}(b) which has an initial position $y^{*}_{i}=0$, $z^{*}_{i}=0.66$ and orientation $\theta_{i}=-30^{0}$. 
Similar to the case of a pusher, the apparent complexity of the trajectory can again be attributed to the fact that it is a superposition of two planar waves of different wavelengths and amplitudes.
But unlike the case of a pusher, here, the characteristics of the planar waves are not determined by the channel dimensions alone, but are also dependent on the initial configuration of the swimmer (similar to Fig.~\ref{fig:2d_neutral}(a)). 
Consequently, slight changes in the initial position or orientation can entirely alter the trajectory of neutral swimmers, as we observed in the case of the square channel. 

%%%%%%%%%%%%%%%%%%%%%%%%%%%%%%%%%%%%%%%%%%%%%%%%%%%%%%%%%%%%%%%%%%%%%%%%%%%%%%%%%%%%%%%%%%%%%%%%%%%%%%%%%%%%%%%%%%%%%
%%%%%%%%%%%%%%%%%%%%%%%%%%%%%%%%%%%%%%%%%%%%%%%%%%%%%%%%%%%%%%%%%%%%%%%%%%%%%%%%%%%%%%%%%%%%%%%%%%%%%%%%%%%%%%%%%%%%%
\section{Conclusions}
\label{sec:conclusion}

In this work, we present a detailed study of the 3-dimensional trajectories of microswimmers inside a micro-channel with square/rectangular cross-section using two methods- (i) full numerical simulations based on lattice-Boltzmann method and (ii) an analytical method based on far-field hydrodynamic approximations combined with method of images. We use a squirmer with tangential surface velocity as a model for the micro-swimmer. The hydrodynamic interactions between the channel walls and the squirmer plays a significant role in defining their 3D trajectories. 

Analysis of the instantaneous velocity of the squirmer at different positions in the channel reveals that the channel confinement and near-wall swimming always decrease the swimming speed of a squirmer when it is oriented parallel to the channel axis, and induce a strong angular velocity.
The far-field analysis captures the nature of these  wall-induced dynamics fairly well till about $90$ \% of the physically possible distance of the microswimmer away from the channel center and towards the walls.
We also provide a detailed description of various two dimensional trajectories produced by the squirmer when it is confined between two parallel plates; which certainly depends on the force/source dipole strengths and the extent of confinement. These observations helped us to underpin the behaviour of swimmers in microchannels with square and rectangular cross sections.
 
In general, our analysis of the trajectories of different squirmers in a 3D channel with square and rectangular cross sections shows that the trajectories are closely dictated by three parameters- (i) strength of the confinement from different directions, (ii) nature and strength of the swimmer, and (iii) initial position and orientation of the swimmer.
Pushers are seen to generally perform helical motion inside a square channel, and a qualitatively similar but more complex trajectory inside a rectangular channel. The far-field calculations predict that for a given force dipole strength of a pusher in a square channel, there is a critical value of the source dipole strength (and hence the size of the swimmer) beyond which the microswimmer only orbits the contours of the channel cross-section without any displacement along the channel axis.
On the other hand, pullers generally make a sliding motion inside the micro-channel.  
Depending on the dipole strength, they choose different paths -- weak pullers slide through the channel center-line, while strong pullers slide through a straight line close to any one of the channel walls.
Neutral squirmers show the most complex 3-dimensional trajectories due to their strong dependency on the initial conditions.
The far-field hydrodynamic calculations, constructed from the  superposition of the hydrodynamic interactions between a squirmer and two pairs of orthogonal walls, can predict the correct trajectories of microswimmers and all the essential features of swimming in two and three dimensions. 
%Furthermore, these calculations can also capture all the essential features of three-dimensional swimming for pullers and pushers in a square channel, except the near-wall sliding of a puller.

Our systematic analysis - determination of instantaneous velocity of the confined swimmer which in turn determine their trajectories, the planar trajectories when the motion of the swimmer is restricted to the symmetric planes of the micro-channel, the three-dimensional trajectories in square channels and finally in the rectangular channels - help to understand the factors that determine the nature of the 3D trajectory of a channel confined micro-swimmer. 
Based on this understanding, we propose a method of superposition to explain the apparent complexity of the 3D trajectories. According to our method, each 3-dimensional trajectory can be constructed from the superposition of two mutually perpendicular planar trajectories (trajectories produced when the squirmer is restricted to move in a plane in the channel). We demonstrate that the superposition method works well to explain the variety of trajectories of pushers, pullers and neutral swimmers in 3D micro-channels.

Hydrodynamics of microorganisms inside different 3D confinements are biologically relevant for understanding their locomotion, survival strategies, and collective behaviour \citep{jeanneret2019confinement,tokarova2021patterns}. 
Clear understanding of the hydrodynamic interactions of microswimmers in strong confinements is also imperative for designing artificial microswimmers and formulating strategies for bio-medical applications, like targeted drug delivery \citep{mou2014autonomous, gao2015artificial}  and targeted interactions with biological cells \citep{gao2013half}. 
However, as we have mentioned in the introduction, a detailed study of the hydrodynamics of microswimmers in 3D confinements remain scarce. 
In this context, the present research work provides a clearer picture of the possible dynamics of different kinds of microswimmers in 3D channels with square/rectangular cross-section.
The unambiguous comparison between numerical simulations and far-field analytical calculations will greatly aid in designing active microfluidic systems for the aforementioned applications in an informed manner.
Furthermore, the proposed superposition method for constructing 3D trajectories of microswimmers in strong confinements can be an efficient way of developing  physical understanding without going into the complexities of full scale numerical simulations of 3D systems and experiments.

\par
\section*{Acknowledgements}
\par
RD acknowledges support from IIT Hyderabad through seed grant no. SG 93. S.P.T. acknowledges the support by Department of Science and Technology, India, via the research grant CRG/2018/000644.
\bibliography{apssamp.bib}% Produces the bibliography via BibTeX.

\end{document}